\documentclass[3p,11pt]{elsarticle}

\usepackage{amssymb}

\usepackage{amsmath}
\usepackage{pstricks}
\usepackage{sgame}
\usepackage{subcaption}
\usepackage{caption}
\usepackage{graphicx}
\usepackage{ulem}
\usepackage{hyperref}
\usepackage{setspace}
\biboptions{numbers,square,comma,sort&compress}

\providecommand{\Ro}{\mathcal{R}_0}

\newcommand{\captionfonts}{\small}
\makeatletter  
\long\def\@makecaption#1#2{%
  \vskip\abovecaptionskip
  \sbox\@tempboxa{{\captionfonts #1: #2}}%
  \ifdim \wd\@tempboxa >\hsize
    {\captionfonts #1: #2\par}
  \else
    \hbox to\hsize{\hfil\box\@tempboxa\hfil}%
  \fi
  \vskip\belowcaptionskip}
\makeatother   

\newcommand{\mastrike}[1]{\textcolor{black}{\sout{#1}}}
\newcommand{\macolor}[1]{\textcolor{black}{#1}}

\newcommand{\micolor}[1]{\textcolor{black}{#1}}

\newcommand{\comment}[1]{}

\journal{arXiv}

\begin{document}

\begin{frontmatter}



\title{Effects of Adaptive Protective Behavior on the Dynamics of Sexually Transmitted Infections}

\author[Epi]{Michael A. L. Hayashi}
\ead{mhayash@umich.edu}
\author[Epi,Math]{Marisa C. Eisenberg}
\ead{marisae@umich.edu}
\address[Epi]{Department of Epidemiology, University of Michigan, Ann Arbor}
\address[Math]{Department of Mathematics, University of Michigan, Ann Arbor}

\begin{abstract}
Sexually transmitted infections (STIs) continue to present a complex and costly challenge to public health programs.  The preferences and social dynamics of a population can have a large impact on the course of an outbreak as well as the effectiveness of interventions intended to influence individual behavior.  In addition, individuals may alter their sexual behavior in response to the presence of STIs, creating a feedback loop between transmission and behavior.  We investigate the consequences of modeling the interaction between STI transmission and prophylactic use with a model that links a Susceptible-Infectious-Susceptible (SIS) system to evolutionary game dynamics that determine the effective contact rate.  The combined model framework allows us to address protective behavior by both infected and susceptible individuals.  Feedback between behavioral adaptation and prevalence creates a wide range of dynamic behaviors in the combined model, including damped and sustained oscillations as well as bistability, depending on the behavioral parameters and disease growth rate.  We found that disease extinction is possible for multiple regions where $\Ro > 1$, due to behavior adaptation driving the epidemic downward, although conversely endemic prevalence for arbitrarily low $\Ro$ is also possible if contact rates are sufficiently high. We also tested how model misspecification might affect disease forecasting and estimation of the model parameters and $\Ro$. We found that alternative models that neglect the behavioral feedback or only consider behavior adaptation by susceptible individuals can potentially yield misleading parameter estimates or omit significant features of the disease trajectory.
\end{abstract}

\end{frontmatter}


\section{Introduction}
\label{sec::introduction}
In spite of advances in treatment and prevention, sexually transmitted infections (STIs) remain endemic worldwide. The CDC estimates that 20 million new cases occur annually in the United States alone \cite{CDC2013}, incurring a total cost of \$16 billion for treatment and care.  Globally, treatable STIs are responsible for approximately 500 million new cases per year \cite{WHO2013}, while an estimated 35 million individuals currently live with HIV. These statistics underscore the importance of understanding the dynamics that drive and sustain STI transmission.  To this end, mathematical epidemiology has made substantial progress investigating the role of contact patterns such as age-structure, sexual networks, and levels of sexual activity \cite{anderson2000mathematical, eames2002modeling}.  However, many open questions remain in understanding the feedback relationship between behavioral change and disease dynamics.  

From a behavioral standpoint, sexually transmitted diseases are noteworthy as they require a direct and intimate interaction between individuals.  As a result, many common preventative measures, such as condom use, are not determined unilaterally \cite{lam2004really, peasant2014condom}.  In addition, assuming individuals form preferences over protective behaviors based on the associated costs and benefits, we would expect these behaviors to respond to the risk of infection as an outbreak progresses \cite{des1996protective, steelfisher2010public, rubin2009public}.  While changes in risky or protective behavior can amplify or suppress outbreaks, adoption of these behaviors can in turn be driven by the spread of disease, as demonstrated by increased testing and condom use among men who have sex with men in response to the HIV outbreak in the US \cite{fisher1992changing, gregson2006hiv}.

Methods from game theory provide a framework with which to capture this feedback, grounded in well-established mathematical and economic theory.  The resulting economic- epidemiological models can explicitly represent the decision process of individuals either in direct interactions (e.g. sexual encounters) or population interactions (e.g. vaccination behavior) \cite{fenichel2011adaptive,schroeder2002game, bauch2005imitation, bauch2004vaccination, reluga2011general, reluga2006evolving, reluga2010game}.  Including the effects of behavioral change on STI dynamics has been primarily motivated by the HIV epidemic among men who have sex with men (MSM) in the 1980s and 1990s, but this modeling approach is relevant for the study of other pathogens and communities as well. Indeed, in 2013 the WHO highlighted the need to study behavioral change in order to design effective interventions \cite{WHO2013}.  

Many economic-epidemiological models rely on two behavioral assumptions that are worth consideration.  The traditional game theoretic framework assumes that all actors are fully rational, responding optimally at every stage of play \cite{fudenberg1991game}.  While convenient, the rationality assumption remains a subject of debate in economic literature.  Empirical studies note circumstances in which individual behavior appears to depart from a strict payoff maximization foundation \cite{selten1990bounded,kahneman1994new,bazerman1983negotiator}.  In addition, most models (e.g.\cite{chen2004rational} and \cite{geoffard1996rational}) assume that only susceptible individuals make choices regarding protective behavior.  This assumption is the result of representing the cost-benefit calculus of individuals as a tradeoff between various private costs of protective behavior and the risk of infection.  However, infected individuals may have incentives to reduce contact or use protection as well, \micolor{motivated by altruism, self interest, or other factors \cite{wolitski2003self, o2005guessing}}.  This has been the focus of several intervention strategies in practice \cite{kennedy2010behavioural, de1998preventive}. 

In this paper, we propose a model of combined behavioral and disease transmission dynamics that uses the outcome of sexual interactions between susceptible and infected individuals to determine the effective contact rate for a mass action model of disease transmission.  \micolor{The combined model bears some similarity to the behavior-disease model proposed in \cite{schroeder2002game}. There are, however, several critical distinctions.  We use a deterministic ODE framework, while our game-theoretic model collapses the protection-use game to a single interaction instead of a multi-stage negotiation.  In addition, similar to \cite{bauch2005imitation} and \cite{reluga2006evolving} we use evolutionary dynamics to represent the process of behavioral change over time, allowing for non-optimal \macolor{but potentially more realistic} behaviors.}  \micolor{Unlike the inductive reasoning game developed by Breban et al. \cite{breban2007mean}, our behavioral dynamics only explicitly consider the current state instead of a history of actions.  While this approach loses some realistic features, it still allows us to relax} the assumption of full rationality while also providing a convenient mathematical formulation for the combined model \cite{smith1982evolution, hofbauer2003evolutionary}.

\section{Model}
\label{sec::methods}
The Susceptible-Infectious-Susceptible (SIS) model has been studied extensively as a simplified representation of bacterial sexually transmitted diseases \cite{anderson1982transmission, anderson1991infectious, yorke1978dynamics}. The model equations are
\begin{equation}
\begin{aligned}
	\dot{S} &= \gamma I - \beta SI, \\
	\dot{I} &= \beta SI - \gamma I,
\end{aligned}
\label{eq::SIS}
\end{equation}
where $S$ is the fraction of the population which is susceptible, $I$ is the fraction infected, \micolor{$\beta$ is the effective contact rate; the product of the rate of sexual partner acquisition and the probability of disease transmission from an infected to a susceptible partner}, and $\gamma$ is the rate of recovery or treatment.  The basic reproductive ratio is
\begin{equation}
	\Ro = \frac{\beta}{\gamma}.
\end{equation}
\micolor{The disease free equilibrium (DFE) occurs if $\Ro < 1$.  Otherwise the endemic prevalence is}
\begin{equation}
	I^* = 1 - \frac{\gamma}{\beta}.
\end{equation}

In order to capture the potential for individuals to adapt their protective behaviors over the course of an outbreak, we define a Bayesian game \cite{harsanyi1967games} between a pair of players\macolor{. In \ref{app::bayesiangames}, we give a brief overview of the definitions and structure of Bayesian games, with more complete treatments given in \cite{fudenberg1991game, tadelis2013game, harsanyi1967games}}.  The payoffs for the game depend on the disease states of both players, which are considered private information.  \micolor{Players must infer the type of their partner, reflecting realistic uncertainty about serostatus \cite{kaplan1993unsafe,gold1996judging,o2005guessing,wiktor1990effect,dawson1994awareness}}.  Consistent with the notation for the SIS model, a player may be one of two types chosen from the type space \micolor{$\Theta = \{S, I\}$}.  Each player chooses between using protection (P) or no protection (U) for a given sexual encounter.  If both players select the same action, the outcome of the game is the same as the chosen action.  We assume that if both players select different actions the encounter does not proceed and the effective contact rate for the pair of players is 0.  
In order to characterize the strategy space for this game, it is convenient to use the type-contingent notation \micolor{$\sigma_{j}(\theta_i) \in \left\{P, U\right\}$} to denote the action player $i$ would choose if she was of type $\theta_i$.  \macolor{A complete strategy for a player then specifies a pair of actions $\sigma_{j} := \sigma_{j}(\theta_i = S)\sigma_{j}(\theta_i = I) \in \left\{PP, PU, UP, UU\right\}$. Without loss of generality, the type-dependent payoff (utility) to player 1 for a given pair of actions and types is written $u_1(\sigma_{j}(\theta_1),\sigma_{k}(\theta_2),\theta_1,\theta_2)$ (where the first entry specifies player 1's action assuming type $\theta_1$ and the second entry player 2's action assuming type $\theta_2$).} \macolor{Then player 1's overall expected payoff for a particular strategy profile (i.e. for a pair of  type-contingent strategies $\sigma_j$ and $\sigma_k$ for each player) is the double expectation of $u_1(\sigma_{j}(\theta_1),\sigma_{k}(\theta_2),\theta_1,\theta_2)$ over both players potential types \cite{fudenberg1991game}, that is,}
\micolor{\begin{equation}
	E[u_1(\sigma_{j}(\theta_1),\sigma_{k}(\theta_2),\theta_1,\theta_2)] = 
	 \sum_{\theta_1} Pr(\theta_1)[\sum_{\theta_2} p_1(\theta_2|\theta_1)u_1(\sigma_{j}(\theta_1),\sigma_{k}(\theta_2),\theta_1,\theta_2)]
\label{eq::expectedutility}
\end{equation}} 
\macolor{where $E[u_1(\sigma_{j}(\theta_1),\sigma_{k}(\theta_2),\theta_1,\theta_2)]$ is sometimes written more simply as $E[u_1(\sigma_{j},\sigma_{k})]$.}
%
The prior probabilities \macolor{for each player being of either type,} \micolor{$Pr(\theta \in \left\{S,I\right\})$}\macolor{,} are given by the distribution of susceptible and infected individuals in the population.  \micolor{We assume that partner selection is not assortative in disease type, so} \macolor{the belief for player 2 being type $\theta_2$ given player 1 being type $\theta_1$ is $p_1(\theta_{2}|\theta_1) = Pr(\theta_{2})$. 
} 
 \macolor{Fig. \ref{fig::payoffs} shows the two possible payoff matrices for the protected sex game that determine the type-dependent payoff terms $u_i(\sigma_{j}(\theta_1),\sigma_{k}(\theta_2),\theta_1,\theta_2)$ in Equation \ref{eq::expectedutility}.  For a concrete example, we compute the expected payoff to player 1 of playing strategy $\sigma_{j} = PU$ if player 2 picks $\sigma_{k} = UU$ when 30\% of the population is infected and we choose $a = 1, b = 0.75, c = 0.5, d = 0.25$.  From Equation \ref{eq::expectedutility} we have }
\begin{equation}
\begin{aligned}
	E[u_1(PU,UU)] &= 0.7(0.7\times0.5 + 0.3\times0.25) \\
	&+ 0.3(0.7\times0.25 + 0.3\times1) \\
	&= 0.44.
\end{aligned} 
\end{equation}
In general, we suppose the specific payoff entries satisfy $a > b > c > d$.  Thus, individuals prefer unprotected sex to protected sex with a partner of the same disease status, preferring either to action pairs resulting in no sexual encounter.  However, individuals prefer protected sex to all other outcomes when their partner is of a different disease type.  This is intended to capture the notion that both susceptible and infected individuals have some incentive to avoid infection or transmission respectively. \micolor{We explore the alternate case where infected individuals do not distinguish between susceptible and infected partners in \ref{app::altcase}}.

\begin{figure}
\centering
\begin{subfigure}[b]{0.45\textwidth}
	\centering
	\begin{game}{2}{2}[Player~1][Player~2]
		     & $P$ & $U$ \\
		 $P$ & $b$ & $c$ \\
		 $U$ & $c$ & $a$ \\
	\end{game}
	\caption{}
	\label{subfig::subgame1}
\end{subfigure}
\begin{subfigure}[b]{0.45\textwidth}
	\centering
	\begin{game}{2}{2}[Player~1][Player~2]
		     & $P$ & $U$ \\
		 $P$ & $b$ & $c$ \\
		 $U$ & $c$ & $d$ \\
	\end{game}
	\caption{}
	\label{subfig::subgame2}
\end{subfigure}
\caption{\macolor{The two variants of the protected sex game for the cases a: $\theta_i = \theta_j$ and b: $\theta_i \neq \theta_j$. }}
\label{fig::payoffs}
\end{figure}

Consequently, for the protected sex game, the time-varying payoff matrix \micolor{$M$} is a $4 \times 4$ square matrix with elements given by Eq. \eqref{eq::expectedutility}.  We can compute the Bayes-Nash equilibrium for any given choice of $a, b, c, d, S, I$.  \macolor{One case of particular interest is at the disease free equilibrium, $I^* = 0$.  Here the game reduces to the $2\times2$ game with payoffs as in Figure \ref{fig::payoffs}a.  This game has two pure strategy Nash equilibria, $(P,P)$ and $(U,U)$. However, the $(U,U)$ equilibrium in the reduced game is both payoff and risk dominant.  Since the $\sigma(\theta_i = I)$ actions do not contribute to the expected payoff, any mixture of the type-contingent strategies $UP$ and $UU$ (similarly $PP$ and $PU$) is a Nash equilibrium in the full model.}  We will revisit this point shortly. 
 
Since we are ultimately interested in the dynamics of behavior at more than a single disease state, we use methods from evolutionary game theory to couple the dynamics of strategy change over time to the disease trajectory. This allows us to close the feedback loop between behavior and disease dynamics. In particular, we use replicator-mutator dynamics with a linear fitness function \cite{hofbauer2003evolutionary, page2002unifying}.  
\macolor{The distribution of strategies in the population using the replicator-mutator approach depends both on the existing distribution of strategies and their expected payoffs, as well as a small degree of random strategy choice. This allows us to capture the notion that individuals may not respond immediately or strictly optimally to the presence of an outbreak, as payoff-suboptimal strategies may remain frequent in the population for some time}. In addition, the possibility of mutation prevents any strategy from becoming extinct.  In the behavioral context, this can be thought of as allowing for a small amount of random choice. \micolor{This feature is particularly important as the replicator equation has stable steady states at strict pure strategy Nash equilibria, \macolor{such that behavior cannot change after fixation on a particular Nash equilibrium even if disease conditions change}.}  \micolor{Both players in the protected sex game have the same action and type sets, so the relative frequency of the $j$th strategy in a large population is $f(\sigma_{j},t)$. }\micolor{The replicator-mutator equation for a given strategy $\sigma_{j}$ is}
\micolor{
\begin{equation}
	\dot{f}(\sigma_{j}) = s[\sum_k f(\sigma_{k},t) q_{kj} \phi(\sigma_{k},t) - \bar{\phi}(t)f(\sigma_{j},t)], 
\end{equation}
}
\micolor{where $q_{kj}$ is the probability that an individual playing $\sigma_{j}$ switches to $\sigma_{k}$, $\phi(\sigma_{j},t) = \sum_k E[u(\sigma_{j},\sigma_{k})]f(\sigma_{k},t)$ is the fitness of $\sigma_{j}$, $\bar{\phi}(t) = \sum_j \phi(\sigma_{j},t)f(\sigma_{j},t)$, and $s \in [0,\infty)$ is a scaling term that determines the speed of behavior change.} \macolor{In vector-matrix form, this can be written as}
\micolor{
\begin{equation}
\dot{\mathbf{f}} = s[D_{\mathbf{f}}Q(M\mathbf{f}) - (\mathbf{f}^T M\mathbf{f})\mathbf{f}],
\label{eq::replicatormutator}
\end{equation}
}
\micolor{where $M$ is the} \macolor{$4\times4$} \micolor{payoff matrix with elements $m_{jk} = E[u_1(\sigma_{j},\sigma_{k})]$, $\mathbf{f} = (f(PP,t), f(PU,t), f(UP,t)$, $f(UU,t))$}, $Q$ is the mutation matrix and \micolor{$D_{\mathbf{f}} = diag(\mathbf{f})$}.    Note that when $Q = I$, the evolutionary dynamics are equivalent to standard replicator dynamics. However, for our simulations we use a mutation probability $\mu = 0.03$ so
\micolor{
\begin{equation}
	Q = (1 - \frac{4}{3}\mu)I + \frac{\mu}{3} \cdot \mathbf{1}.
\end{equation}
}
In a large population, the aggregate effective contact rate is determined by the average outcome over all pairs, so $\beta = \frac{\beta_b S_U I_U}{SI}$ where $\beta_b$ is the baseline effective contact rate, \micolor{$S_U = S\times(f(UP,t)+f(UU,t))$} is the proportion of susceptible individuals playing $U$, and \micolor{$I_U = I\times(f(PU,t)+f(UU,t))$} is the proportion of infected individuals playing $U$.  

The combined model equations with evolutionary behavioral dynamics can be written as
\micolor{\begin{equation}
\begin{aligned}
	\dot{S} &= \gamma I - \beta_b S_U I_U, \\
	\dot{I} &= \beta_b S_U I_U - \gamma I, \\
	\dot{\mathbf{f}} &= s[D_{\mathbf{f}}Q(M\mathbf{f}) - (\mathbf{f}^T M\mathbf{f})\mathbf{f}].
\end{aligned}
\end{equation}}
\micolor{The system above has six compartments.  Of these, only four are strictly necessary as $S + I = 1$ and $\sum_j f(\sigma_{j},t) = 1$.}   \micolor{In a completely susceptible population, action pairs where both susceptibles play $P$ or $U$ are Bayes-Nash equilibria. In order to restrict the domain of possible initial conditions, we take the risk dominant equilibria, or mixtures between $UP$ and $UU$.  This can be interpreted as individuals in a disease free state preferring unprotected sex because it best hedges against uncertainty in partner actions.} \micolor{While the $\sigma(\theta_i = I)$ actions are never realized at the DFE, the type-contingent strategy framework implies that individuals must be able to specify an action that they would take if they became infected.}  The effective contact rate for a single infected individual is \micolor{$(f(UU,0)+f(PU,0))\beta_b$}, where \micolor{$f(UU,0),f(PU,0) \in [0,1]$}.  As a result, the basic reproductive rate of the combined model is
\micolor{\begin{equation}
	\Ro = \frac{(f(UU,0)+f(PU,0)) \beta_b}{\gamma}.
\label{eq::R0}
\end{equation}}
\micolor{Thus, $\Ro$ depends on} \macolor{the behavioral} \micolor{initial condition, unlike the standard SIS model.  The value of the above expression will lie within the interval $[0, \beta_b / \gamma]$.} \macolor{The payoff and risk dominance of the $(U,U)$ Nash equilibrium at the initial disease-free steady state (discussed above) suggests that $f(PU,0)$ is likely to be low, and indeed for the remainder of this paper we will assume that the initial infected plays the strategy $UU$, while the underlying susceptible population at $t=0$ will be assumed to take on a mixture of the $UU$ and $UP$ strategies. It is nonetheless of interest to note that the dependence of $\Ro$ on the initial behavioral conditions suggests the potential for both high $\Ro$ with disease extinction and low $\Ro$ with endemic prevalence: if $f(PU,0) \approx 1$ it may be possible to have high $\Ro$ but still have disease extinction due to all susceptibles playing the protected strategy, while conversely if $\Ro$ is low due to low frequency of $UU$ and $PU$ but otherwise has a high contact rate, it may be possible to generate endemic prevalence even for $\Ro < 1$ (as explored further below).  }

\begin{figure}
	\centering
	\begin{subfigure}[b]{0.475\textwidth}
		\includegraphics[width=\textwidth]{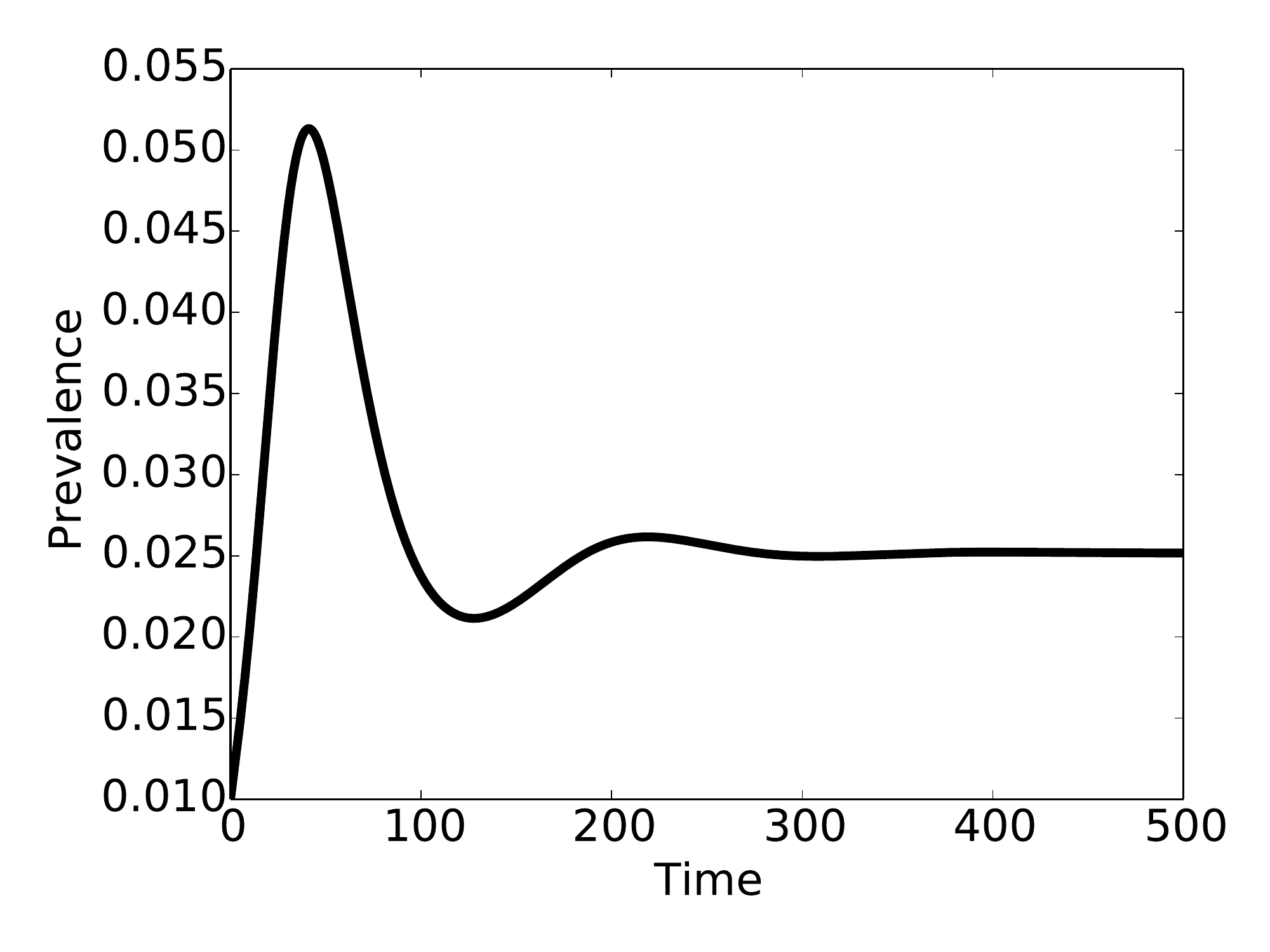}
		\caption{}
	\end{subfigure}
	\begin{subfigure}[b]{0.475\textwidth}
		\includegraphics[width=\textwidth]{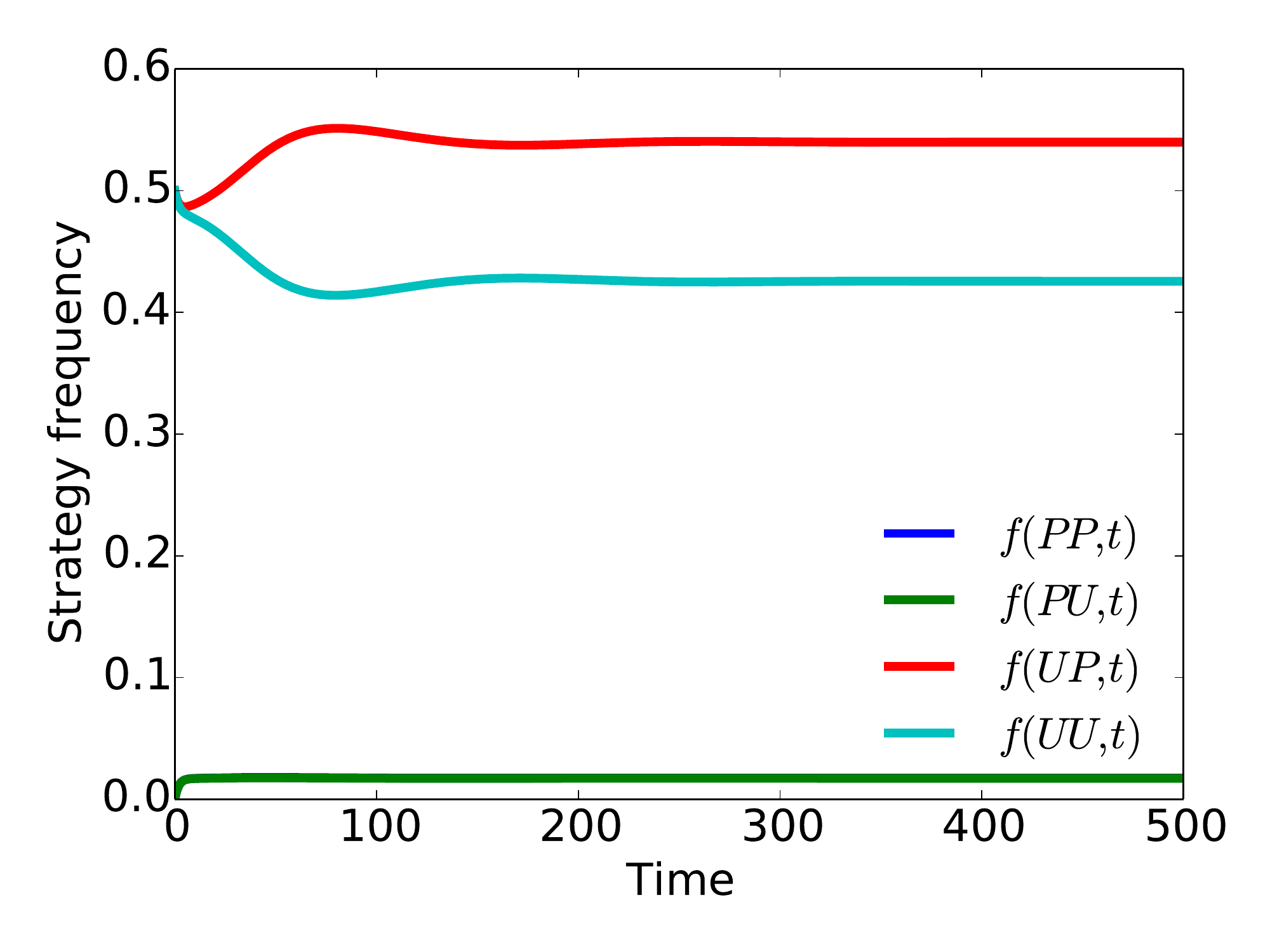}
		\caption{}
	\end{subfigure}
	\begin{subfigure}[b]{0.475\textwidth}
		\includegraphics[width=\textwidth]{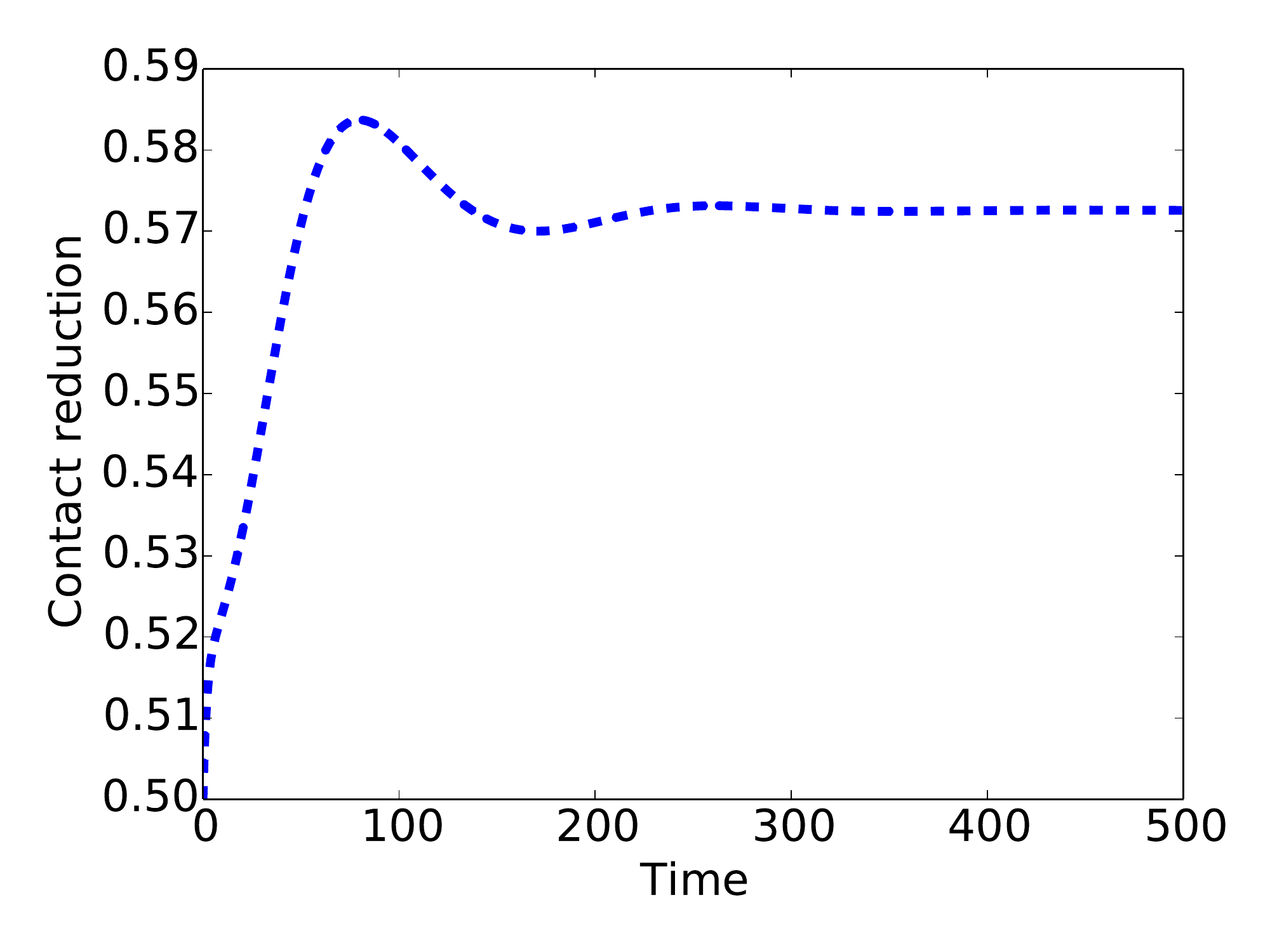}
		\caption{}
	\end{subfigure}
	\begin{subfigure}[b]{0.475\textwidth}
		\includegraphics[width=\textwidth]{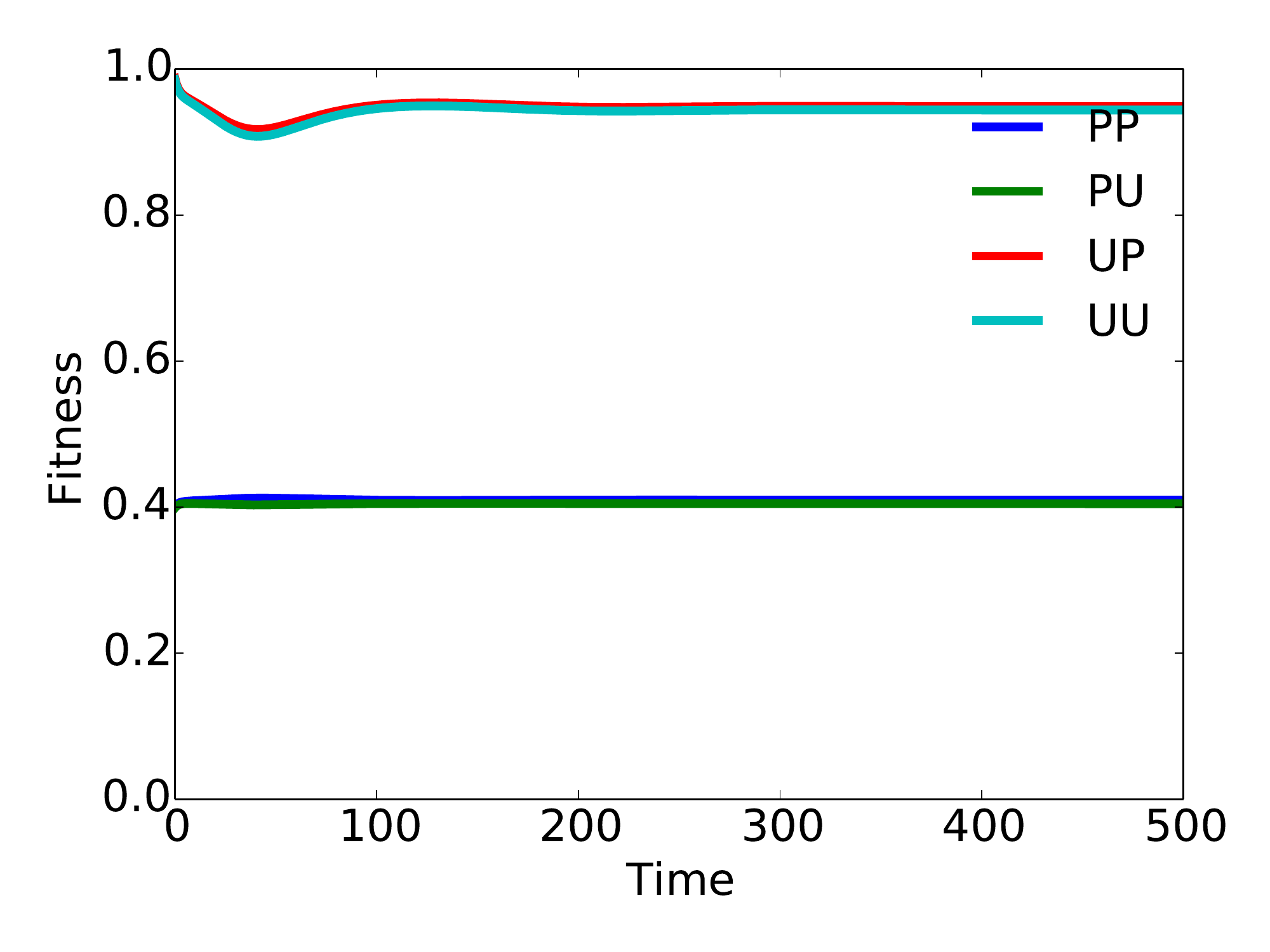}
		\caption{}
	\end{subfigure}
	\caption{Simulated trajectories for the combined model with $R_0 = 1.2$.  Disease parameters \micolor{$\beta_b = 1.2, \gamma = 0.5$} with behavioral parameters $a = 1, b = 0.6, c = 0.4, d = 0.2$.  (a) Disease prevalence, (b) Type-contingent strategies, (c) Fractional contact reduction, (d) Strategy fitness.}
	\label{fig::trajectories}
\end{figure}

\section{Results}
\label{sec::results}

Due to the addition of the replicator-mutator equations, it is not straightforward to solve for the steady states of the combined model analytically.  Instead, we conducted a range of numerical simulations to investigate the effects of behavioral dynamics using Python 2.7 with Numpy 1.9, Scipy 0.15.0, and Matplotlib 1.4.2.  \micolor{In order to explore a range of outcomes generated by the combined model, we focused on both long-term and short-term dynamics.  Given the addition of behavior change, the short term dynamics provided useful mechanistic insights to explain the steady state properties of the model.} Unless noted otherwise we used $a = 1, b = 0.6, c = 0.4, d = 0.2$ for the values of the outcome payoffs. \micolor{In general the qualitative features of the model do not change substantially for different payoff values provided the overall scale and ordering are preserved.} \macolor{Unless otherwise specified, the disease initial conditions were $S(0) = 0.99$, $I(0) = 0.01$. In the first section below, we consider fixed initial behavioral conditions at $f(UU,0) = f(UP,0) = 0.5$, while in the following section we consider the effects of varying the initial behavioral conditions (for $f(UU,0)$ between 0 and 1).} 

\subsection{Model Dynamics} We began by inspecting an example set of model trajectories, shown in Fig. \ref{fig::trajectories} (with $\beta_b = 1.2, \gamma = 0.5,$ and \micolor{$f(UU,0) = 0.5$} to give $\Ro = 1.2$), to illustrate the interactions between behavioral changes and disease dynamics.  Comparing the prevalence and contact reduction trajectories demonstrated the impact of the adaptive behavioral dynamics on the progression of the simulated outbreak.  In this example the outbreak was relatively small, so susceptible individuals did not have a large incentive to use protection.  However, the initial increase in disease prevalence favored strategies where infected individuals used protection, which in turn decreased the force of infection enough to halt the initial outbreak.  However, as prevalence decreased, unprotected strategies became less costly and contact reduction decreased again.  Consequently there was a small secondary outbreak.  The timing of the behavioral response also appeared to be influential.  While contact reduction tracked prevalence over time, it did so at a delay.  Thus the level of contact reduction was still relatively high at the onset of the second outbreak, preventing a large secondary peak.  This preliminary exploration suggested that both the infectiousness of the disease and the speed of behavioral adaptation play important roles in the overall model dynamics.  

\begin{figure}
	\centering
	\begin{subfigure}[b]{0.475\textwidth}
		\includegraphics[width=\textwidth]{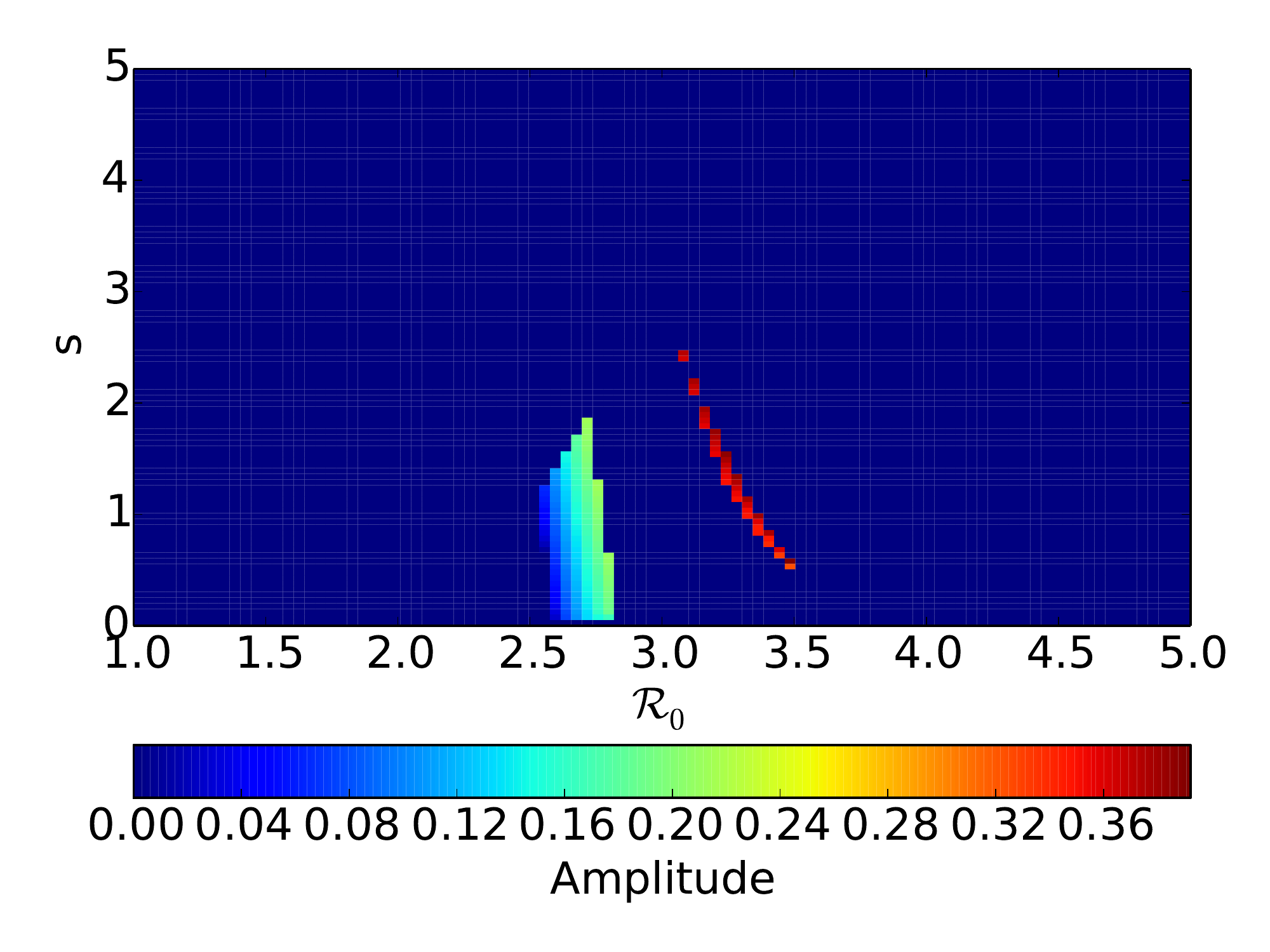}
		\caption{}
	\end{subfigure}
	\begin{subfigure}[b]{0.475\textwidth}
		\includegraphics[width=\textwidth]{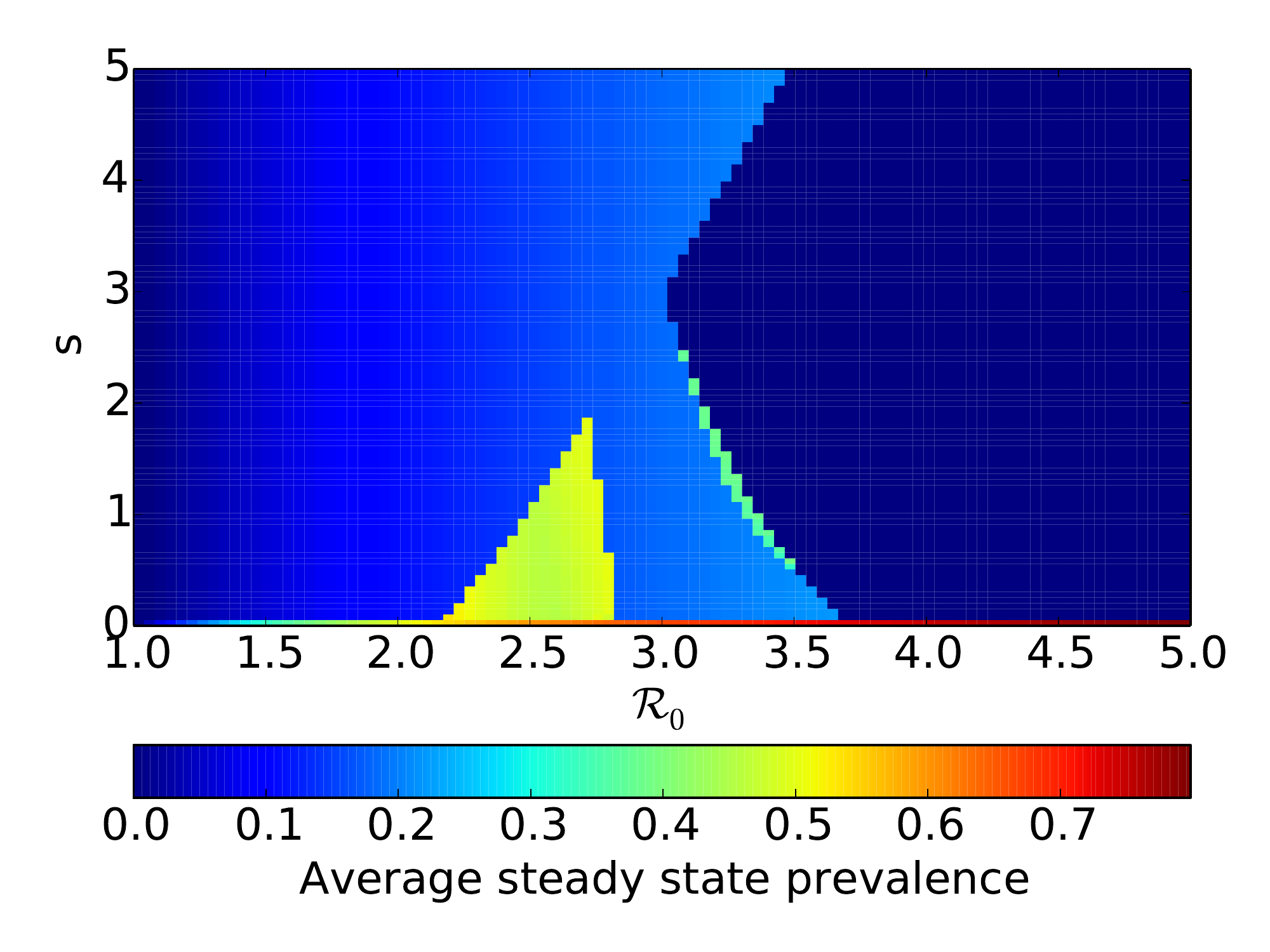}
		\caption{}
	\end{subfigure}
	\caption{\mastrike{D} \macolor{Long term d}ynamics of the combined model with increasing values of $\Ro$ and the behavioral scale parameter $s$. $\Ro$ was adjusted by varying the parameter \micolor{$\beta_b$} in Eq. \eqref{eq::R0}. (a) The amplitude of steady state prevalence oscillations.  (b) The average prevalence at steady state.}
	\label{fig::bifurcation}
\end{figure}

To explore this issue further, we defined a timescale parameter $s$ for the behavioral dynamics, given as a scaling factor on the payoff matrix $A$ in Eq. \eqref{eq::replicatormutator}. From a behavioral perspective, this parameter can be thought of as controlling the speed of adaptation in the population. We then evaluated the long-term dynamics of the model in steady state for a range of values of $\Ro$ and $s$, shown in Fig. \ref{fig::bifurcation} (with all remaining parameters as in Fig. \ref{fig::trajectories}).  In the triangular region about $\Ro = 2.7$ and the canyon between $3 < \Ro < 3.5$ the damped oscillations became sustained with a substantially higher average steady state prevalence. The triangular region of oscillations gives way to another endemic steady state region as $\Ro$ increases, followed by the thin band of oscillations and eventually disease extinction in the triangular region along the right edge.

\begin{figure}
	\centering
	\begin{subfigure}[b]{0.475\textwidth}
		\includegraphics[width=\textwidth]{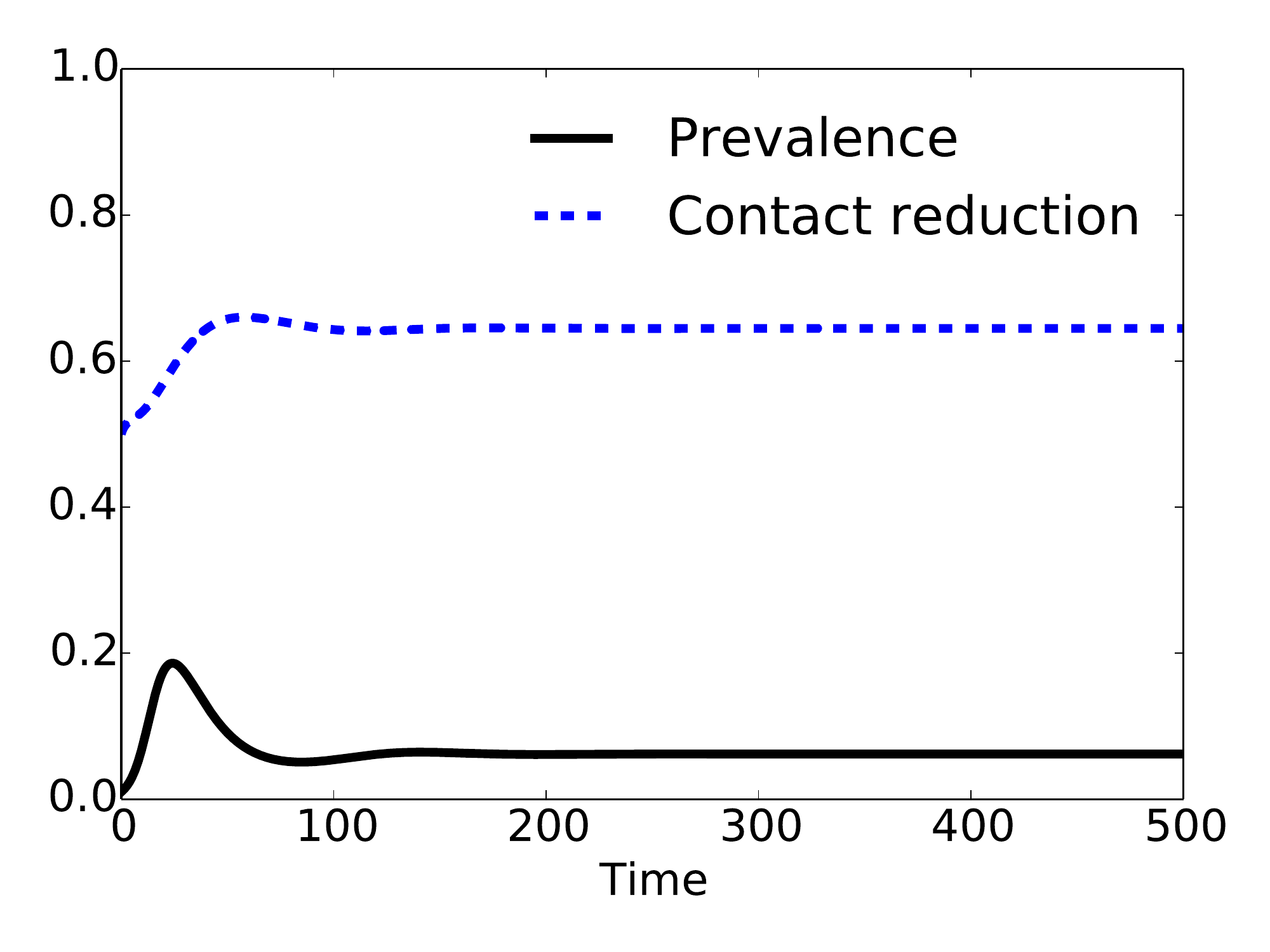}
		\caption{}
	\end{subfigure}
	\begin{subfigure}[b]{0.475\textwidth}
		\includegraphics[width=\textwidth]{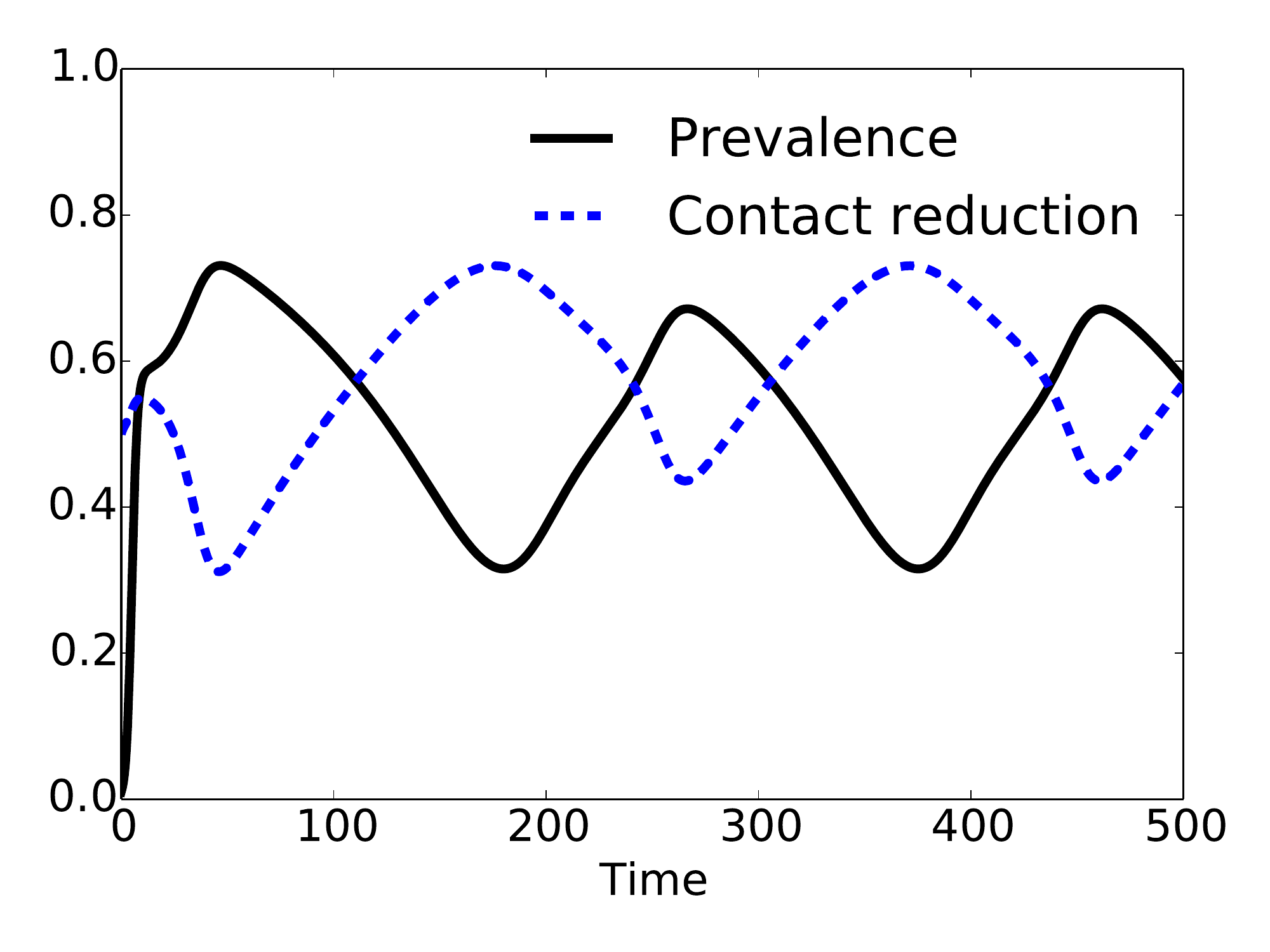}
		\caption{}
	\end{subfigure}
	\begin{subfigure}[b]{0.475\textwidth}
		\includegraphics[width=\textwidth]{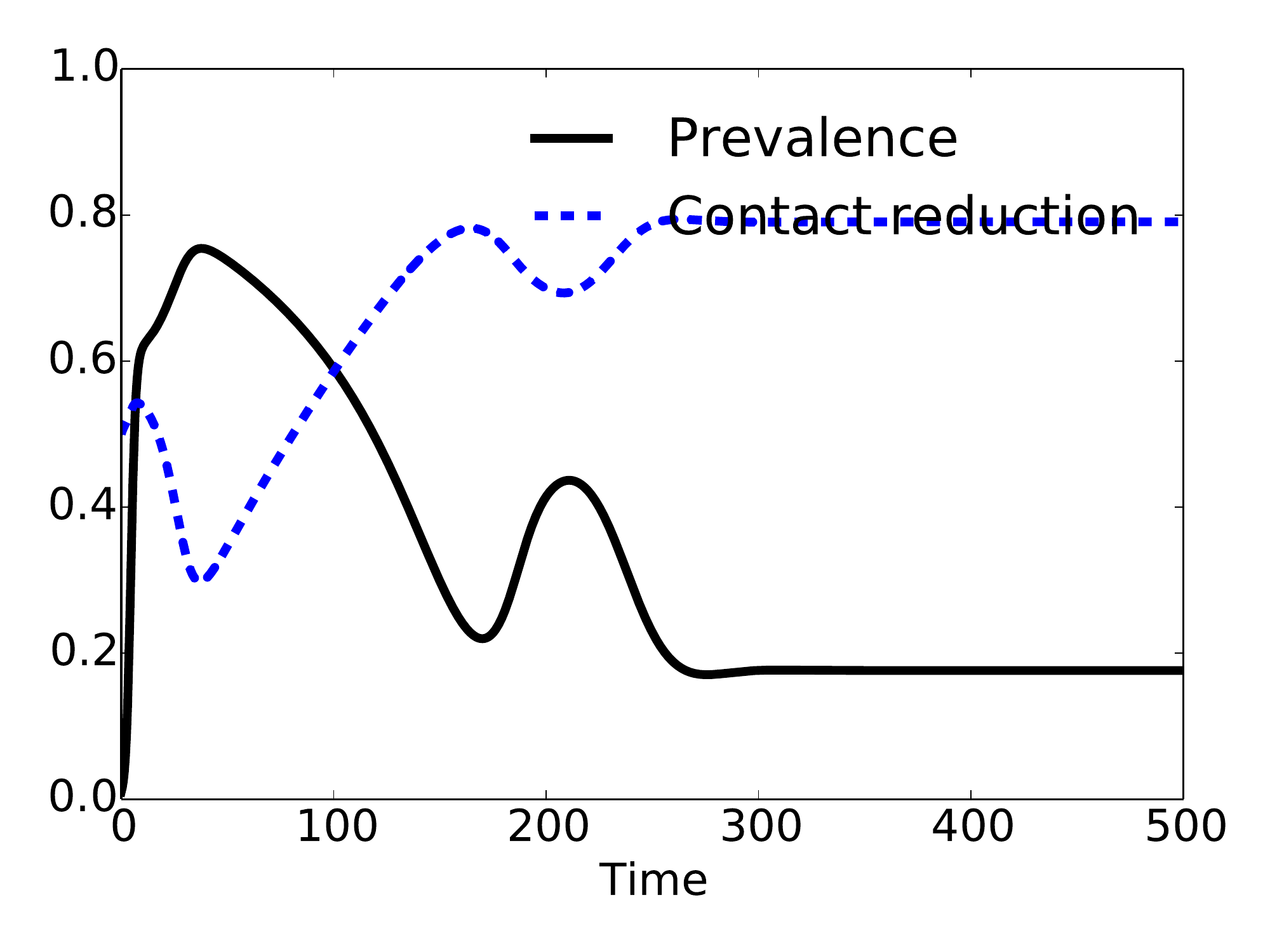}
		\caption{}
	\end{subfigure}
	\begin{subfigure}[b]{0.475\textwidth}
		\includegraphics[width=\textwidth]{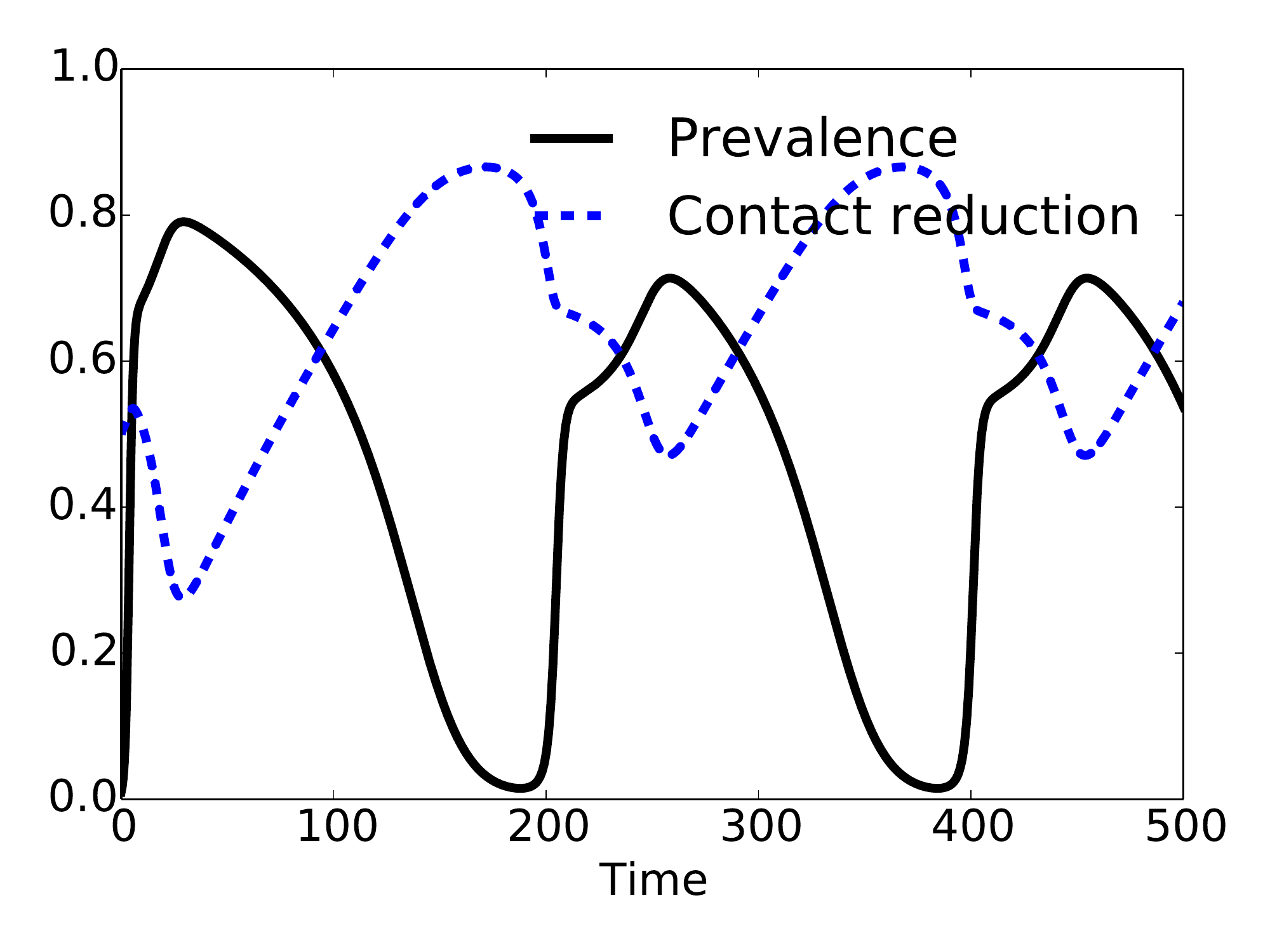}
		\caption{}
	\end{subfigure}
	\begin{subfigure}[b]{0.475\textwidth}
		\includegraphics[width=\textwidth]{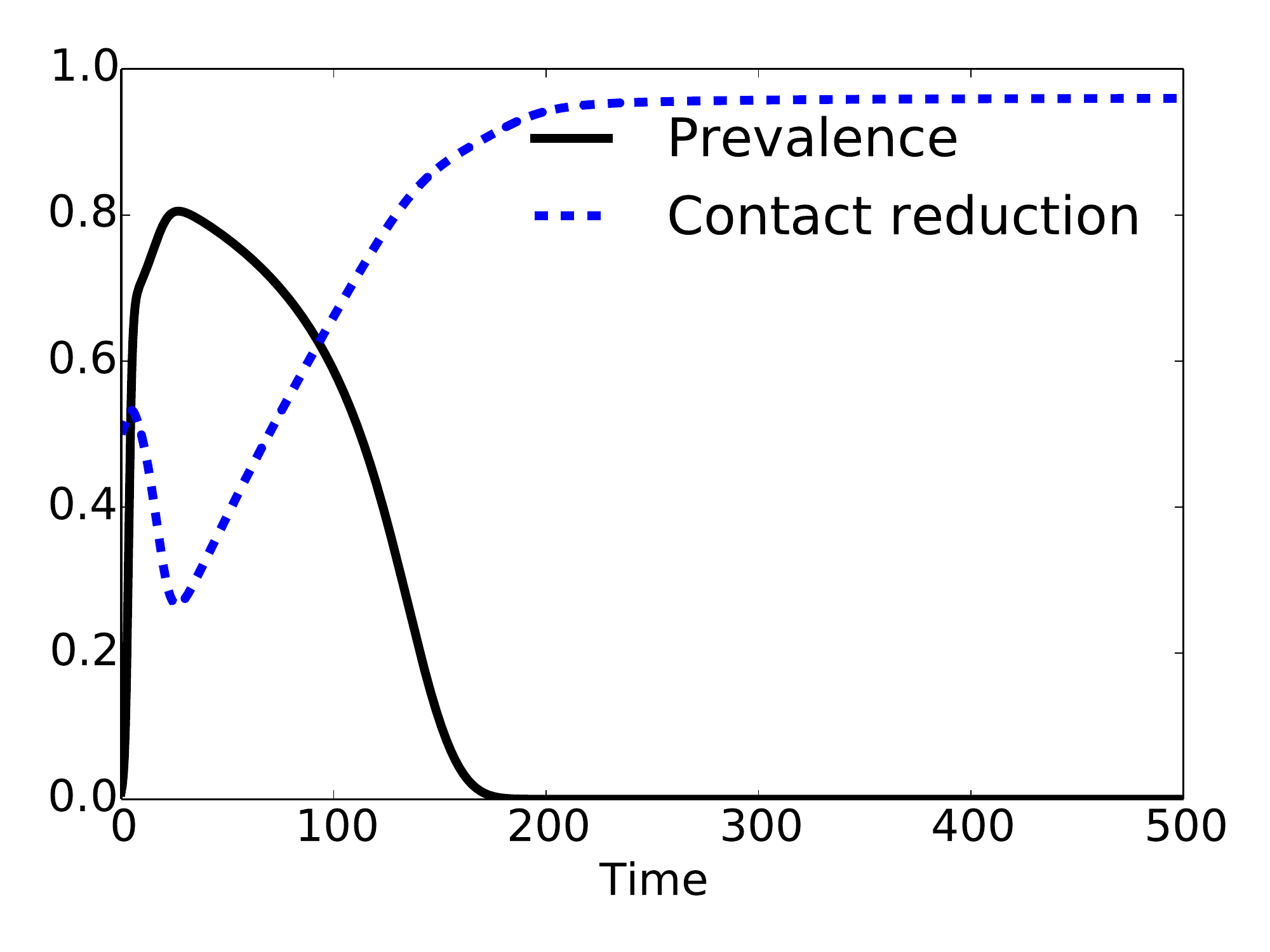}
		\caption{}
	\end{subfigure}
	\caption{Prevalence and contact reduction from the combined model with \micolor{$\gamma = 0.5, \mathbf{f}(0) = (0.0,0.0,0.5,0.5)$ for all simulations and (a) $\beta_b = \Ro = 1.5$, (b) $\Ro = 2.7$, (c) $\Ro = 2.9$, (d) $\Ro = 3.3$, and (e) $\Ro = 3.5$.}}
	\label{fig::manytrajectories}
\end{figure}

\begin{figure}
	\centering
	\includegraphics[width=0.475\textwidth]{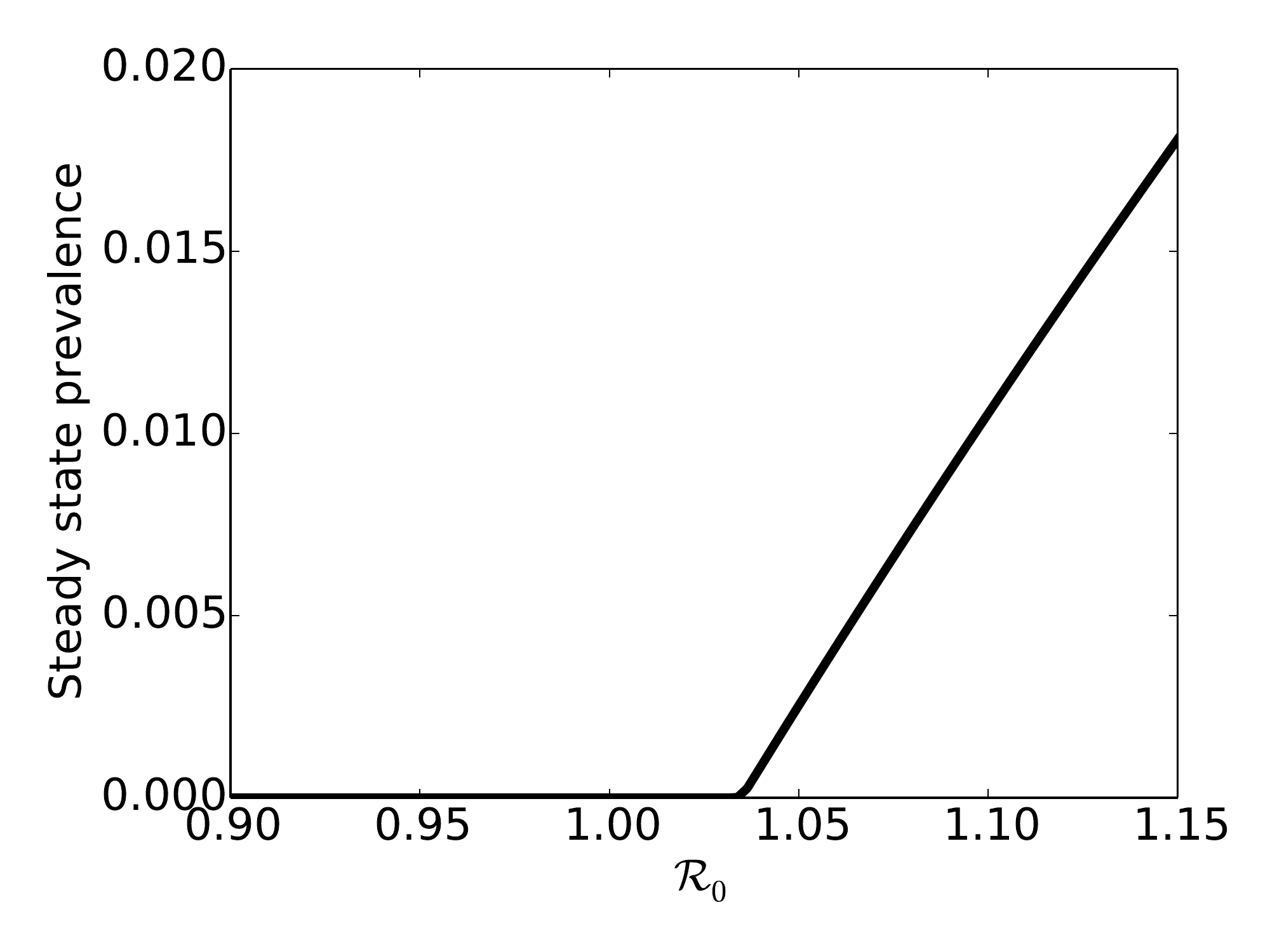}
	\caption{Average steady state prevalence of the combined model for increasing $\Ro$ with \micolor{$\gamma = 0.5, \mathbf{f}(0) = (0.0,0.0,0.5,0.5)$}.  The transition between the disease free and endemic equilibrium occurs at $1 < \Ro < 1.05$. }
	\label{fig::herdimmunity}
\end{figure}

Figure \ref{fig::manytrajectories} and \macolor{Supplementary Figure} \ref{S1_Fig} provide a more detailed examination of the dynamics in Fig. \ref{fig::bifurcation} as $\Ro$ increases for $s = 1$.  For $\Ro$ sufficiently above 2, the early increase in contact reduction by infected individuals was not sufficient to stop the spread of the initial outbreak (\ref{fig::manytrajectories}b).  As a result prevalence increased, causing infecteds to begin to switch back to unprotected strategies (\ref{S1_Fig} b).  However, at higher prevalence susceptible individuals had an incentive to use protection, compensating for the behavior of the infected population.  As in the initial example, this overall increase in contact reduction reduced the force of infection and halted the growth of the outbreak.  The magnitude of this response appeared to determine the long term dynamics of the model.  When prevalence was reduced to a moderate level neither protective nor unprotected strategies were able to gain a stable foothold in the population, leading to sustained oscillations in both contact reduction and prevalence (\ref{fig::manytrajectories}b, \ref{S1_Fig} b).  A stronger contact reduction response pushed the disease to a level where infected individuals again had a strong incentive to use protection.  The fitness advantage of this strategy was sufficient to survive the second outbreak, leading to an endemic steady state (\ref{fig::manytrajectories}c, \ref{S1_Fig} c).  Notably, when $\Ro$ was high enough that the resulting behavioral response nearly drove the disease to extinction, sustained oscillations were again possible (\ref{fig::manytrajectories}d, \ref{S1_Fig} d).  Subsequent outbreaks in this scenario were contained by increasing contact reduction from a higher baseline due to the initial outbreak, but grew rapidly due to the infectiousness of the disease.  Finally, when the initial outbreak was extremely large, essentially all remaining susceptibles \micolor{(including those who recently recovered) played} protective strategies, driving the disease to extinction (\ref{fig::manytrajectories}e, \ref{S1_Fig} e). Since the boundaries of both oscillatory regions shifted with both $s$ and $\Ro$, it appeared that the timescales of infection and adaptation must align in order to produce sustained oscillations (e.g. Figure \ref{fig::manytrajectories}b and \ref{S1_Fig}b). \macolor{In particular, oscillatory dynamics appear only to be possible for relatively slow adaptation speeds, suggesting that the lag induced in the behavioral dynamics contributes to the potential for oscillations.}

Using the same parameters as in Fig. \ref{fig::manytrajectories}, in Figure \ref{fig::herdimmunity} we also examined steady state prevalence near $\Ro = 1$, to evaluate whether behavior changes may affect the threshold for generating an outbreak. We found that steady state prevalence remained zero even when $\Ro$ was greater than 1, suggesting that behavioral dynamics were able to extinguish the disease even when the initial growth rate would have generated and epidemic.

\begin{figure}
	\centering
	\begin{subfigure}[b]{0.475\textwidth}
		\includegraphics[width=\textwidth]{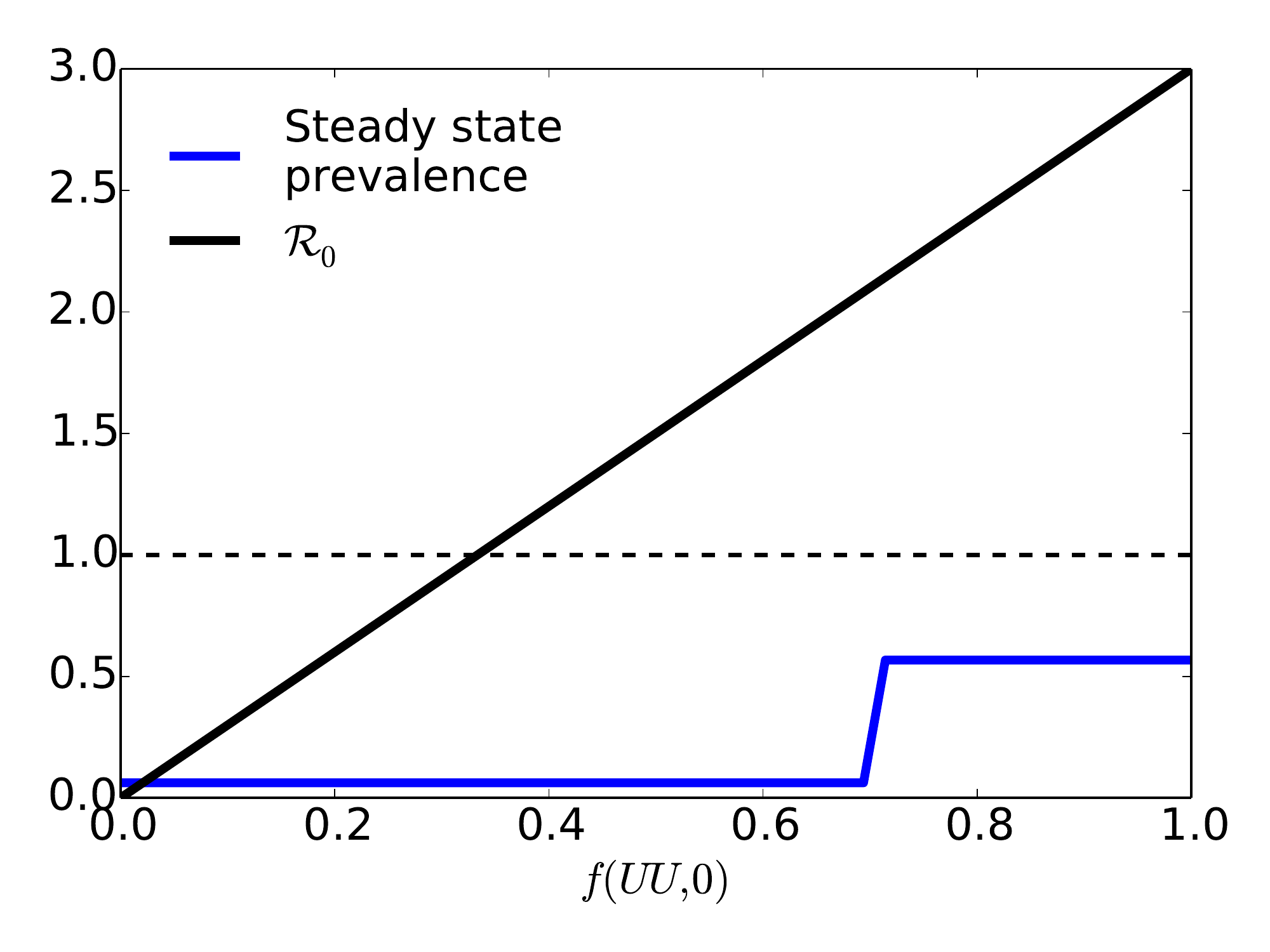}
		\caption{}
	\end{subfigure}
	\begin{subfigure}[b]{0.475\textwidth}
		\includegraphics[width=\textwidth]{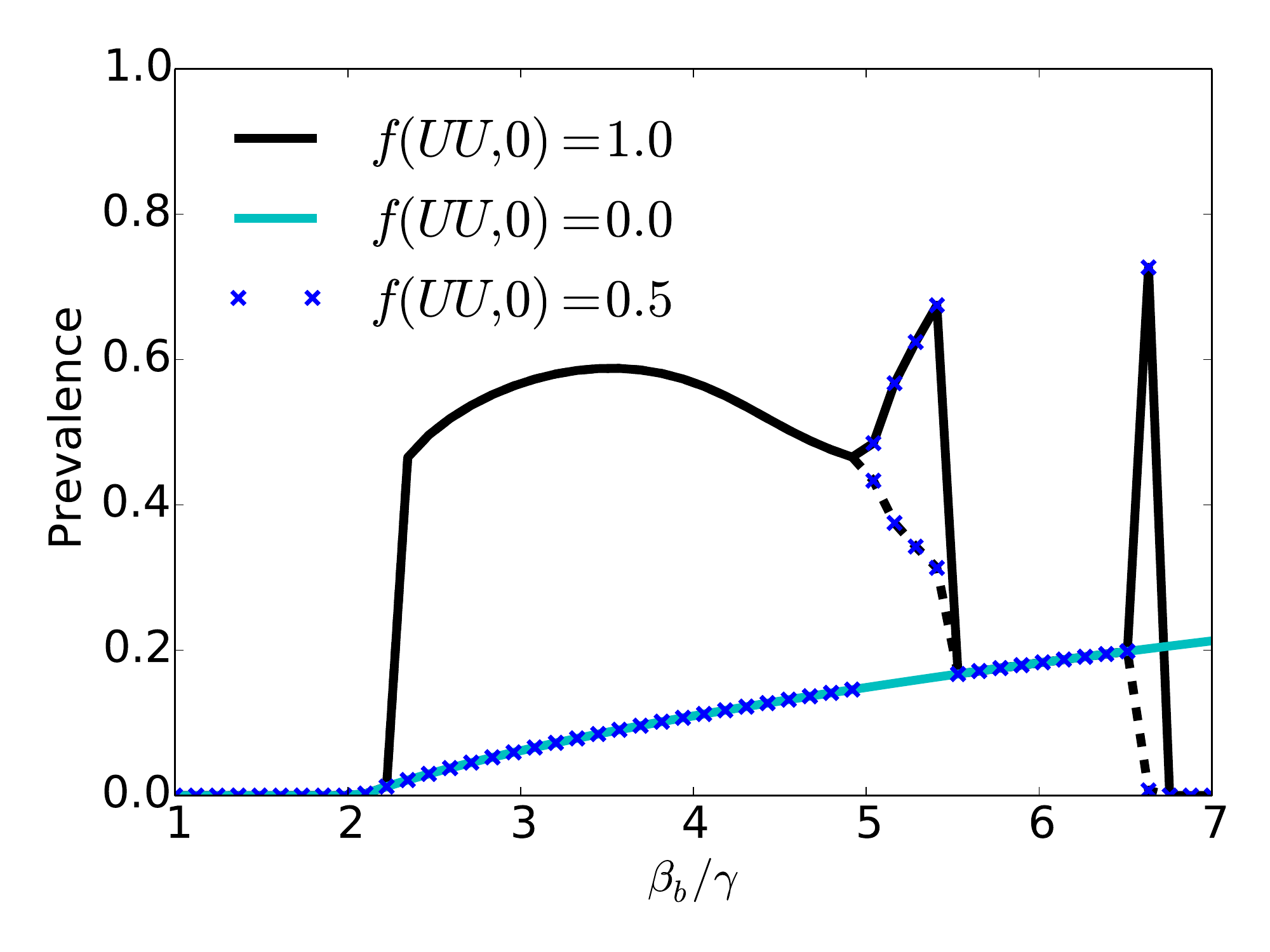}
		\caption{}
	\end{subfigure}
	\caption{The effect of changing the initial condition $f(UU,0)$ with \micolor{$\gamma = 0.5$}.  (a) Steady state prevalence and $\Ro$ for $\beta_b/\gamma = 3$ (Dashed line indicates where $\Ro = 1$). (b) Steady state prevalence of the combined model as $\beta_b / \gamma$ increases for three choices of $f(UU,0)$.  In regions of $\beta_b / \gamma$ where sustained oscillations occurred, the solid black line shows the value of the crest while the dashed line shows the value of  the trough.}
	\label{fig::UUinitial}
\end{figure}

\subsection{Behavioral Initial Condition Analysis} For our next set of simulations, we tested the effect of varying the initial behavioral conditions.  Figure \ref{fig::UUinitial}a shows the average steady state prevalence across $f(UU,0) \in [0,1]$ when $\beta_b / \gamma = 3$ \macolor{(where the remaining population was assumed to take strategy $UP$ as noted above). As suggested by the form of $\Ro$, a}n endemic steady state persisted even for $\Ro < 1$.  This counterintuitive phenomenon occurred because, while the disease initially declined, it did not immediately become extinct (Fig. \ref{S2_Fig}).  The declining prevalence led to increased adoption of $UU$, driving the effective reproductive rate back above one. Figure \ref{fig::UUinitial}b depicts the effect of different levels of $f(UU,0)$ on the steady state behavior of the combined model for increasing $\beta_b / \gamma$. A wide range of dynamics were observed as $\beta_b/\gamma$ varied, including a single stable equilibrium, bistability, and oscillations. When $f(UU,0)$ was low (green line in Fig. \ref{fig::UUinitial}b), the model only exhibited damped oscillations, and did not produce the same extinction behavior at high $\beta_b/\gamma$ as in the case where $f(UU,0)$ was substantially larger than zero.  For $f(UU,0)=0.5$ (blue dotted line), the model exhibited the same bifurcation pattern as in the previous section.  At high $f(UU,0)$ (black line) the behavior was similar, however the endemic prevalence before the first oscillatory region was substantially greater. From this, $f(UU,0)$ appeared to act as a switch between possible steady state regimes of long term dynamics.

For the constant steady state regions, the endemic prevalence took one of two values for a given $\beta_b / \gamma$ depending on $f(UU,0)$, similar to Fig. \ref{fig::UUinitial}a. More broadly, for any fixed $\beta_b / \gamma$, there were two basins of attraction corresponding to either the $f(UU,0) = 0$ or $f(UU,0) = 1$ case, where the unstable equilibrium dividing the two regions depended on the value of $\beta_b / \gamma$. This is illustrated by the $f(UU,0) = 0.5$ curve in Fig. \ref{fig::UUinitial}b, which switches between existing in the $f(UU,0) = 0$ and $f(UU,0) = 1$ basins around $\beta_b/\gamma = 5$. There are also regions of Fig. \ref{fig::UUinitial}b where only a single steady state exists regardless of the value of $f(UU,0)$, at very low values of $\beta_b/\gamma$ (left portion just above $\beta_b/\gamma = 2$) and larger values (between the oscillatory regions). 

As noted in the previous section, there were also multiple regions where the long term dynamics showed disease extinction even though $\Ro >1$. For $f(UU,0) = 0.5$ and $1$, the left corner of Fig. \ref{fig::UUinitial}b shows  extinction with $\Ro > 1$, and the right corner shows the same extinction discussed in the previous section for the $f(UU,0) = 1$ basin at high values of $\beta_b/\gamma$.

\subsection{Comparison with Alternate Models} 
\begin{figure}
	\centering
	\begin{subfigure}[b]{0.475\textwidth}
		\includegraphics[width=\textwidth]{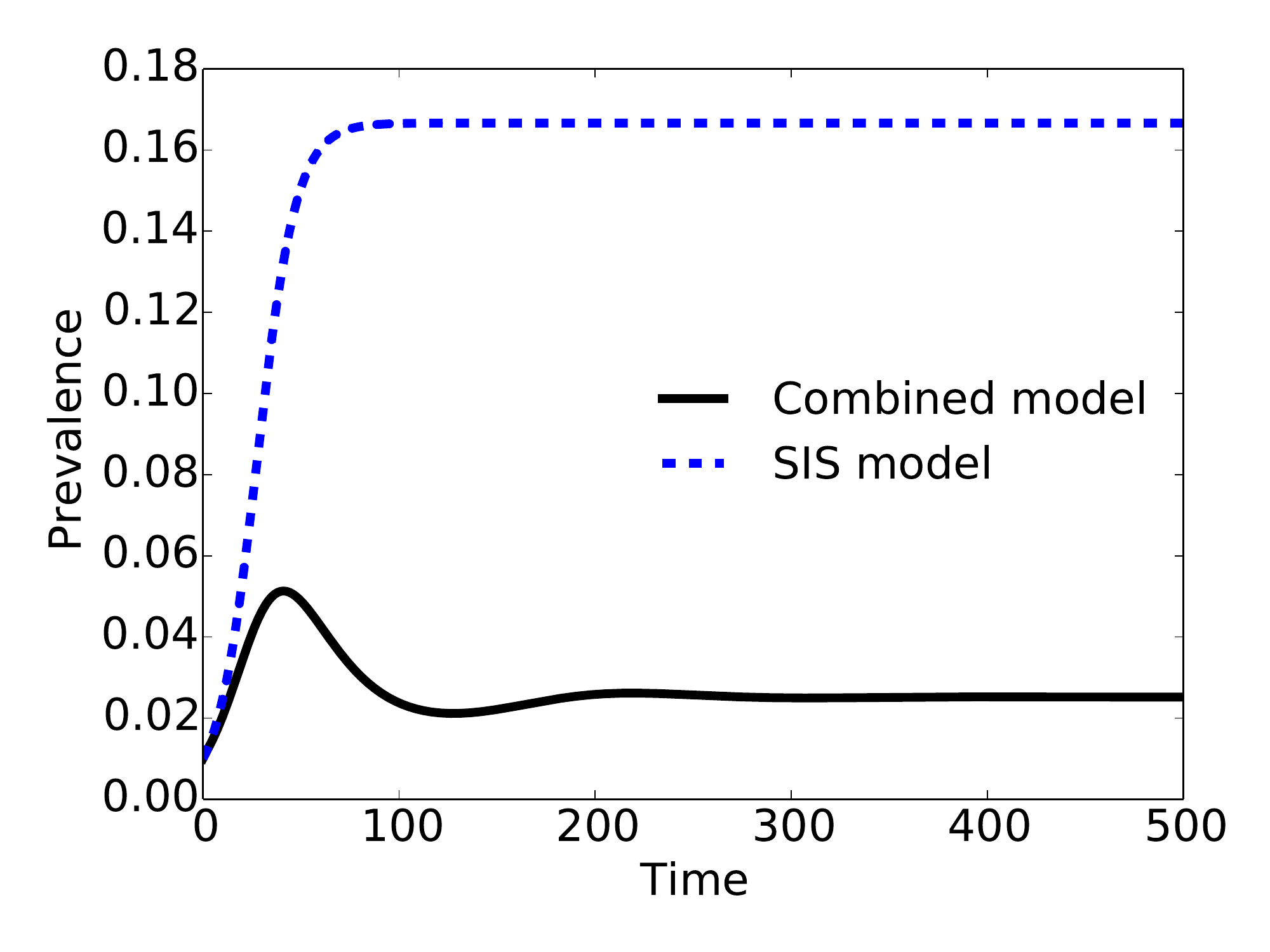}
		\caption{}	
	\end{subfigure}
	\begin{subfigure}[b]{0.475\textwidth}
		\includegraphics[width=\textwidth]{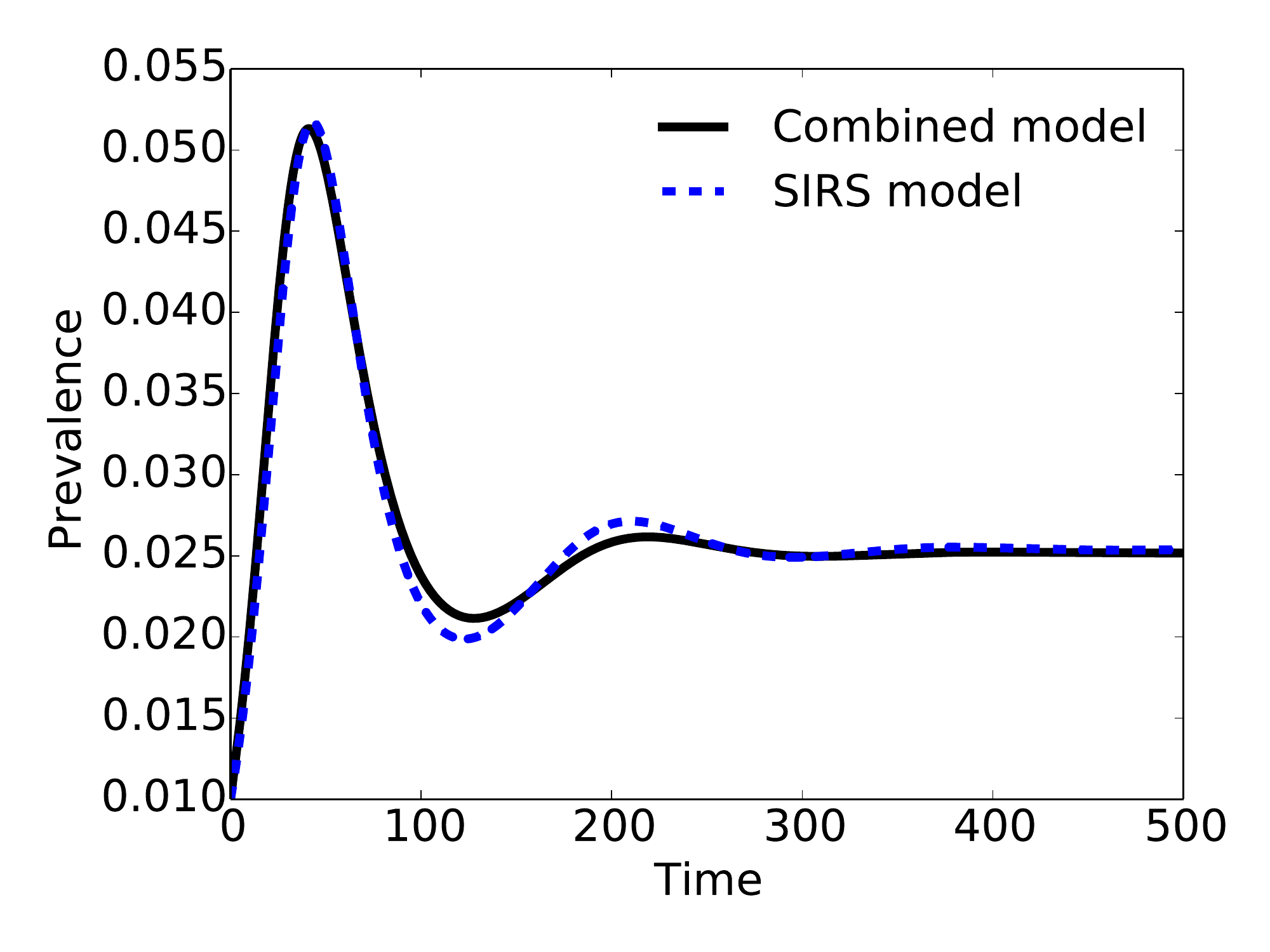}	
		\caption{}
	\end{subfigure}
	\begin{subfigure}[b]{0.475\textwidth}
		\includegraphics[width=\textwidth]{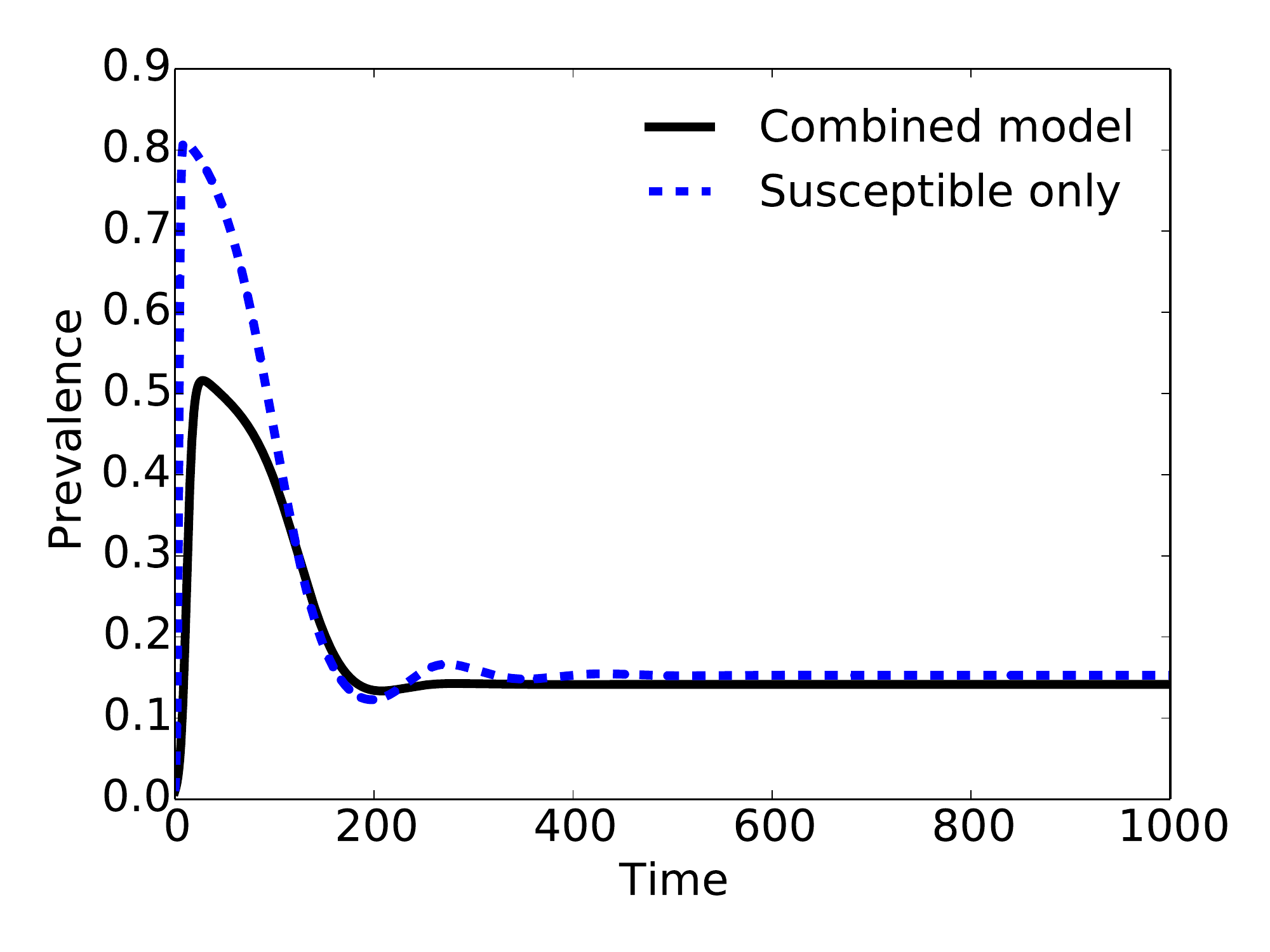}	
		\caption{}
	\end{subfigure}
	\caption{Comparison of simulations of the combined model other compartmental models. (a) SIS model contact rate \micolor{$\beta = 0.6, \gamma = 0.5$} to match the $R_0 = 1.2$ of the combined model, (b) Prevalence trajectories for the combined model and an SIRS model fit to the combined model prevalence simulated with $\Ro = 1.2$.  SIRS best fit parameters: \micolor{$\beta = 0.274, \gamma = 0.203, \delta = 0.022$}. \micolor{(c) A variant combined model in which only susceptible individuals adapt their behavior fit to the simulated trajectory from the full model with $\beta_b = 1.8, \gamma =0.5$ and $\Ro = 2.7$.  The best fit parameter $\beta_b^* = 2.84$ for the reduced model, giving an $\Ro = 5.68$.}}
	\label{fig::comparison}
\end{figure}

Finally, we considered how the dynamics of the model compare to other potential models of the disease dynamics, to examine how neglecting the behavioral dynamics may alter model forecasts of the epidemic trajectory or affect estimates of key epidemiological parameters such as $\Ro$. As a first example, we chose a fixed contact rate SIS model parameterized such that the $\Ro$ of both models was identical.  While the trajectories were similar at very early times (at which the epidemic growth rate can still be characterized by $\Ro$), the combined model quickly diverged due to the reduced effective contact rate, and equilibrated at a substantially lower endemic level (Figure \ref{fig::comparison}a).  Similarly, we computed $\Ro = \frac{1}{1 - I_{\infty}}$ naively from the steady state prevalence of the combined model in Figure \ref{fig::comparison}a, without accounting for a time-varying contact rate.  The estimated value of 1.03 was substantially lower than the true $\Ro$ of 1.2 for the combined model. This difference can be even more severe when considering endemic steady states at higher $\Ro$'s (such as the area to the right of the triangular region in Figure \ref{fig::bifurcation}). For example, the endemic steady state in Figure \ref{fig::manytrajectories}c yielded an apparent $\Ro$ of 1.21, when the underlying $\Ro$ for the combined model was 2.9. Given the simplicity of the SIS model, these discrepancies were not surprising.  \macolor{Nonetheless}, they highlight some pitfalls of neglecting the effects of behavioral dynamics.

However, as the SIS model is known not to be able to produce oscillations, it might not be a likely choice given data that came from real-world infection dynamics similar to the combined model. Due to the delay between recovering from infection and returning to susceptibility, the SIRS model is capable of producing damped oscillations, and so might make a reasonable initial guess for the model structure if the length of immunity was unknown.  To evaluate how a more realistic but misspecified model might affect parameter estimation, we fit the SIRS model (\ref{app::altmodels}) to the prevalence trajectory of the combined model with $\Ro = 1.2$, using least squares with Nelder-Mead optimization, as shown in Figure \ref{fig::comparison}b.  While the best fit SIRS model conformed well to the target trajectory, the resulting parameter estimates included both a four fold decrease in the baseline contact rate and a substantial increase in the average waiting time until a recovered individual becomes susceptible again.  The $\Ro$ for the best fit SIRS model was 1.35, a 12.3\% increase compared to the behavioral model. In this case, the feedback between behavior and transmission may make interventions more effective than would appear to be the case if we had assumed a fixed contact rate.

Finally, most previous work combining game theory with transmission models to model contact reduction has focused on the adaptive behavior of susceptibles \cite{geoffard1996rational, chen2004rational, reluga2010game, bauch2005imitation}.  This is a natural formulation for vaccination, but may be less suitable for more general contact reduction behaviors.  Thus, to test the effect of modeling adaptive behavior by both susceptible and infected individuals, we used a reduced variant of our combined model in which infected individuals always select the action $U$.  \micolor{As with the SIRS model, we fit this reduced model to a simulated trajectory from the full model.  In this case, the reduced model overshoots the initial outbreak curve, but equilibrates to a similar endemic equilibrium to the full model (Figure \ref{fig::comparison}c).  However, the best fit parameter and $\Ro$ for the reduced model were nearly two times higher than the true values for the full model.  For a wider range of parameter values of the full model, the reduced model generally failed to provide qualitatively accurate fits.  In addition, the reduced model did not produce the same range of dynamics as the full model.  We give a full description of the reduced model in \ref{app::altmodels} as well as an expanded set of fits and simulations.}

\section{Discussion}
\label{sec::discussion}

To explore the feedback between behavior and disease dynamics, we developed a model of disease transmission with adaptive behavior among both susceptible and infected individuals.  Numerical simulations illustrated the effect of behavior-disease feedback on model dynamics and inferences about STI transmission (Figure \ref{fig::trajectories}).  We found that phase transitions between damped and sustained oscillations occurred at an intermediate transmission rate and again at a higher transmission rate (Figure \ref{fig::bifurcation}).  Increasing the adaptation rate parameter $s$ reduced the range of transmission rates at which sustained oscillations could occur.  

Disease extinction occurred both for low and high transmission rates (Figures \ref{fig::bifurcation}, \ref{fig::UUinitial}).  At lower transmission rates, adoption of protection by infected individuals lead to extinction above the typical $\Ro = 1$ threshold (Figure \ref{fig::herdimmunity}).  This suggests that behavior-disease feedback can create a herd-protection-like effect if the infected population can be reduced sufficiently to incentivize the use of protection among that group.  At high transmission rates and low-to-mid range levels of initial risky behavior ($f(UU,0)$) in the infected population, we observed the counterintuitive result that a large initial increase in risky behavior ($UU$ and $PU$ strategies) among infected individuals \macolor{preceded} disease extinction.  \micolor{However, because susceptible individuals were also switching to protective strategies \macolor{($PP$ and $PU$)} rapidly, \macolor{the combined effect of both susceptibles and infected adopting $PU$ resulted in sufficient growth of} $PU$ to drive the disease to extinction as the effective number of susceptible individuals remained small \macolor{and all were largely playing protective strategies} (Figure \ref{fig::manytrajectories}, \ref{S1_Fig}).  In fact, both \macolor{the high $\Ro$ and low $\Ro$} extinction phenomena \macolor{(for $\Ro>1$ but still low)} appear to be driven by a reduction in the effective size of the `bottleneck' population group.}  \macolor{The low $\Ro$ extinction (Figure \ref{fig::herdimmunity}) seems largely due to increasing infected protective strategies, while the high R0 extinction seen on the right side of Figure \ref{fig::manytrajectories} is largely due to increasing susceptible protective strategies. In each case, the low/high prevalence makes infecteds/susceptibles a bottleneck group, where sufficient protective response in that group appears to result in disease elimination. The feedback between disease and behavior thus} often constrains endemic outcomes to scenarios where a disease is only moderately infectious, for a range of realistic levels of initial risky behavior (Figure \ref{fig::UUinitial}b).

However, simulations predicted a more pessimistic outcome for changes in the initial strategic distribution.  In general, the initial level at which infected individuals used protection did not determine whether the disease would become extinct.  As long as the value of $\beta_b / \gamma$ was sufficiently large, any $f(UU,0)$ led to an outbreak and endemic disease (Figure \ref{fig::UUinitial}).  Instead, the initial distribution acted as a switch between two steady state regimes: one in which initial risky behavior by infecteds (high $f(UU,0)$) lead to sustained oscillations and higher endemic prevalence could occur for values of $c_b$ below the first oscillatory region, and another in which initial protective behavior by infecteds (low $f(UU,0)$) yielded damped oscillations and generally lower endemic prevalence.  The exception to this pattern is the region where $\beta_b / \gamma$ is large.  As noted above, in this region extinction occurs when early risky behavior by infected individuals is then countered by widespread use of protection by susceptible individuals.  However, early widespread protection use by infecteds leads to a smaller outbreak that is slowed, but not eliminated by susceptible protection use.  \macolor{This complicates the interpretation of $\Ro$ for the combined model, as the standard endemic threshold property often does not apply.} For interventions it is thus important to understand the interaction between early behavioral patterns, the transmission rate, and treatment levels. Interventions based on these factors may yield counterintuitive results depending on the other factors, making comprehensive data collection critical when optimizing intervention strategies. 

\micolor{It is also interesting to note that while an analysis of the Bayes-Nash equilibria of the protected sex game at DFE indicates that the susceptible-unprotected strategies \macolor{$(U,U)$} are risk dominant, our combined model suggests that the actual initial distribution will depend on a population's previous experience with disease.  That is, when the outbreak is large (i.e. for high $\Ro$\macolor{, Figure \ref{fig::manytrajectories}}), adoption of protective strategies result\macolor{s in disease elimination. However, the increased protective strategies by susceptibles mean that the model switches to the attraction basin of the non-dominant $(P,P)$ Nash equilibrium as it goes to the new disease-free steady state \ref{S1_Fig} (or as close as possible given the mutation rates in the replicator-mutator equation)}.  By contrast, below the threshold for high $Ro$ extinction, the model tends to reach a steady state close to the susceptible-unprotected \macolor{$(U,U)$ Nash equilibrium} since endemic prevalence is low enough to favor susceptible-susceptible contact. \macolor{This suggests that in general, the disease-free strategy balance between protected and unprotected depends on the magnitude of previous outbreaks, with larger outbreaks yielding disease-free protective behaviors even though they are neither payoff nor risk dominant.}  As a result, post-outbreak surveillance and control are crucial to prevent recurrent outbreaks due to reintroduction.}

Based on comparisons between the combined model and similar disease models with simplified contact dynamics, 
predictions from fixed-contact rate models may omit important dynamical features or yield misleading parameter estimates (Figure \ref{fig::comparison}). Similarly, if an STI model is misspecified, the common method of computing $\Ro$ from endemic prevalence when a disease is assumed to be at equilibrium \cite{heffernan2005perspectives} is not appropriate and yields inaccurate estimates. One advantage of using a game theoretic framework to model interactive behavior is the flexibility of ordinal utility.  Since protective measures can be used by susceptible or infected individuals to prevent infection or transmission, respectively, we used a preference order that gave individuals in both disease states an incentive to use protection if paired with a partner of the opposite state.  Our simulation results suggest that disease dynamics differ in this scenario as opposed to when only susceptible individuals adapt their behavior\macolor{, and indeed model fits using the reduced model considering only susceptible behavior change often yielded both incorrect $\Ro$ estimates and dynamic trajectories (Figure \ref{S4_Fig2})}.  The role of infected behavior dynamics was particularly notable in the early stage of outbreaks when disease was not prevalent enough to induce susceptible individuals to use protection.  In addition, the shift between infected and susceptible adoption of protective strategies appeared to drive sustained oscillations. \macolor{To more thoroughly consider the effects of infected behavior dynamics, we also considered an alternative preference structure in which infected individuals no longer explicitly prefer protected sex with susceptible individuals, but rather have the same preferences regardless of their partner's disease status (\ref{app::altcase}). This model only exhibits one oscillatory region and a higher endemic prevalence than the original model (Fig. \ref{S5_Fig}). However, extinction still occurs at high $\Ro$, confirming that susceptible protective strategies drive the high-$\Ro$ extinction phenomenon.}

In order to focus on the effect of adaptive protective behavior, our combined model simplifies many other realistic factors that contribute to STI transmission.  \micolor{The protected sex game does not include explicit negotiation, which likely biases the effective contact rate downward, since negotiation or coercion could lead to unprotected sex even when the initial action pair is $(U,P)$.}  The combined model presented here still uses mass action assumptions to determine interactions between individuals.  Although more complex contact patterns are known to influence transmission dynamics, we model a homogeneous population to focus on the effect of time-varying behavior. While this may not be completely implausible in a highly active \micolor{group such as MSM frequenting bathhouses,} it is almost certainly a poor representation of the manner in which individuals form sexual partnerships.  \micolor{Typically, adding contact heterogeneity increases the $\Ro$ of a model, so we might expect more rapid early outbreak growth for a wider range of parameters in our model, potentially expanding the regions in which sustained oscillations or even disease extinction occur.  Our general framework, however, is amenable to extensions to expand the state space of the disease model to represent more complex natural histories or population structures as with standard transmission models.}  In addition, it is possible to model the combined process on a contact network or using a stochastic framework.  \micolor{One particular extension that could yield insight into spatial patterns and the effect of local information would be a simulation of the model on a regular lattice, where individuals could use either a global or local prior to estimate the probability of their partner's type.}

A  particularly compelling consequence of developments in economic-epidemiological models is the potential to estimate more complex behavioral parameters using traditional surveillance data. In the context of our combined model, this would result in estimates for the relative payoff values $(a, b, c, d)$, which capture useful information about the perceptions of populations facing STI outbreaks.  While subject to as yet unknown identifiability properties, this manner of parameter estimation could provide a valuable link between game theoretic methods and the extensive empirical literature on the epidemiology of STIs.

\section*{Acknowledgements}
This work was partially supported by a University of Michigan Block Grant to MALH.

\section*{References}
\bibliographystyle{elsarticle-num}
\bibliography{game_theory_std_hayashi_eisenberg}

\pagebreak
\appendix
\section{\micolor{Bayesian Games}}
\label{app::bayesiangames}
\micolor{This section provides a brief overview of static Bayesian games of incomplete information.  Those desiring a more complete treatment may refer to \cite{fudenberg1991game},\cite{tadelis2013game},and \cite{harsanyi1967games}.}  

\micolor{
\subsection{Definition}
}
\noindent\micolor{A normal form $n$-player symmetric static Bayesian game includes}
\micolor{
\begin{itemize}
	\item A set of players $N = \left\{1,2,\ldots,n\right\}$.
	\item Actions $a_i \in A$ for each player.  
	\item Types $\theta_i \in \Theta$ for each player.
	\item Pure type-contingent strategies $\sigma: \Theta \to A$.  By convention, we use the notation \\$\sigma_{j} = \sigma_{j}(\theta_i^1)\sigma_{j}(\theta_i^2)...\sigma_{j}(\theta_i^m)$ to represent the $j$th strategy for a finite type space.
	\item Belief distributions $p_i$ where $p_i(\theta_{-i}|\theta_{i})$ is the conditional distribution on the types of other players given player $i$'s type.  $-i$ denotes the set of players except i.
	\item Type-dependent payoffs $u_i: A^n \times \Theta^n \to \mathbb{R}$ for each player.  The expected type-dependent payoff ($E[u_i(a_1,...,a_n|\theta_i)]$) gives the average payoff over player i's belief regarding the other players' types conditional on player i's own type.  $E(u_i(a_1,...,a_n,\theta_1,...,\theta_n))$ is player $i$'s unconditional average payoff over all $n$ player types, sometimes denoted $E(u_i(\sigma_{j_1},...,\sigma_{j_n}))$ in terms of strategies.
\end{itemize}
}
\macolor{In general, a \textit{profile} of types, actions, or strategies is defined to be a listing of the particular types/actions/strategies (respectively) assigned to each player.} \micolor{While the actions in a static Bayesian game take place simultaneously, it is useful to break the game down into stages as follows}
\micolor{
\begin{enumerate}
	\item Nature chooses a profile of types $(\theta_1,\theta_2,\ldots,\theta_n)$ from the common prior distribution.
	\item Each player picks a strategy $\sigma_{j_i}$
	\item Each player learns only his type $\theta_i$.
	\item Using Bayes' theorem and the common prior, each player forms beliefs $p_i(\theta_{-i}|\theta_i)$ over other players' types.   
	\item Players choose actions simultaneously according to their strategy ($a_1 = \sigma_{j_1}(\theta_1)$) to form a profile $(a_1,a_2,\ldots,a_n)$.
	\item Players receive their payoffs $u_i(a_1,...,a_n,\theta_1,...,\theta_n)$ based on the action profile and the type profile.
\end{enumerate}
}

\micolor{
\subsection{Bayesian Nash Equilibrium}
}

\micolor{The strategy profile $(\sigma_{j_1}^*,\ldots,\sigma_{j_n}^*)$ is a Bayesian Nash equilibrium if for all players and all types, $\sigma_{j_i}^*$ satisfies}
\micolor{
\begin{equation}
	\sum_{\theta_{-i}} p_i(\theta_{-i}|\theta_i)u_i(\sigma_{j_i}^*(\theta_{i}),\sigma^*_{j_{-i}}(\theta_{-i}),\theta_i,theta_{-i}) \geq \sum_{\theta_{-i}} p_i(\theta_{-i}|\theta_i)u_i(\sigma_{k_i}(\theta_i),\sigma^*_{j_{-i}}(\theta_{-i}),\theta_i,\theta_{-i})
\end{equation}}
\micolor{for any $\sigma_{k_i} \neq \sigma^*_{j_i}$.  Equivalently
\begin{equation}
	\sum_{\theta_i} Pr(\theta_i)[\sum_{\theta_{-i}} p_i(\theta_{-i}|\theta_i)u_i(\sigma_{j_i}^*(\theta_{i}),\sigma^*_{j_{-i}}(\theta_{-i}),\theta_i,\theta_{-i})] \geq \sum_{\theta_i} Pr(\theta_i)[\sum_{\theta_{-i}} p_i(\theta_{-i}|\theta_i)u_i(\sigma_{k_i}(\theta_i),\sigma^*_{j_{-i}}(\theta_{-i}),\theta_i,\theta_{-i})]
\end{equation}}

\pagebreak

\section{Behavioral Trajectories for Simulations in Figure \ref{fig::manytrajectories}} 
\begin{figure}[h!]
	\centering
	\begin{subfigure}[b]{0.475\textwidth}
		\includegraphics[width=\textwidth]{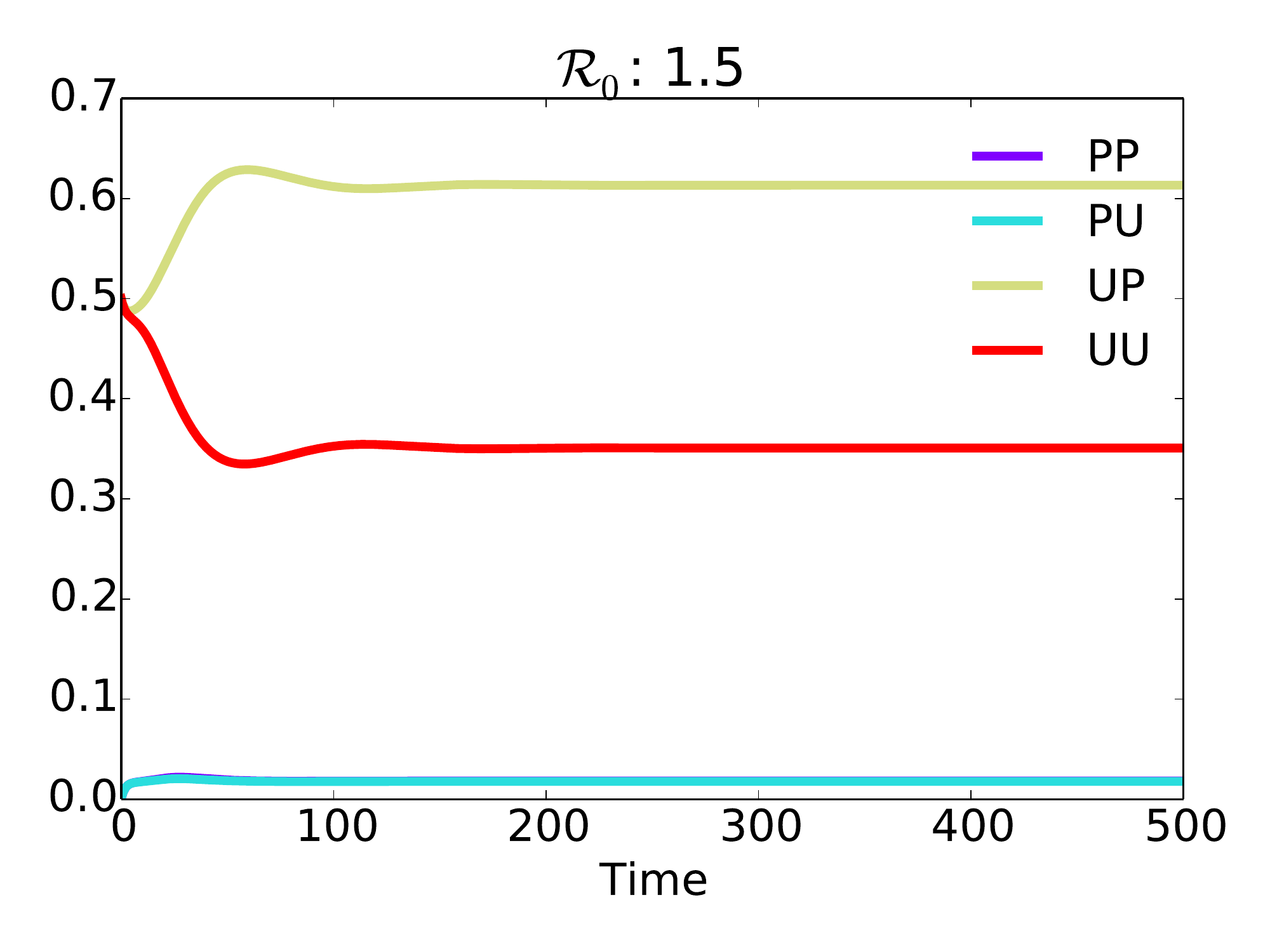}
		\caption{}	
	\end{subfigure}
	\begin{subfigure}[b]{0.475\textwidth}
		\includegraphics[width=\textwidth]{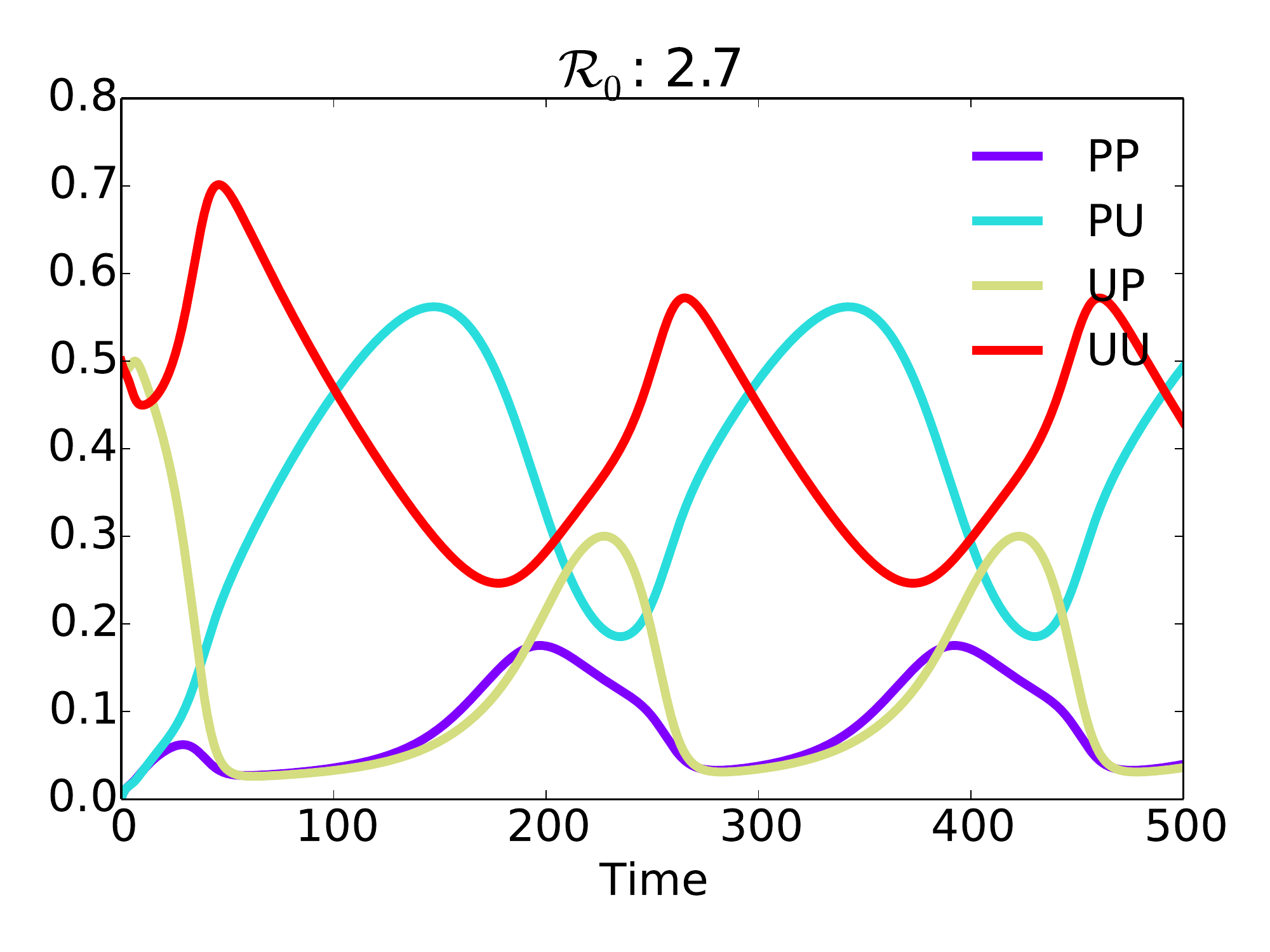}
		\caption{}	
	\end{subfigure}
	\begin{subfigure}[b]{0.475\textwidth}
		\includegraphics[width=\textwidth]{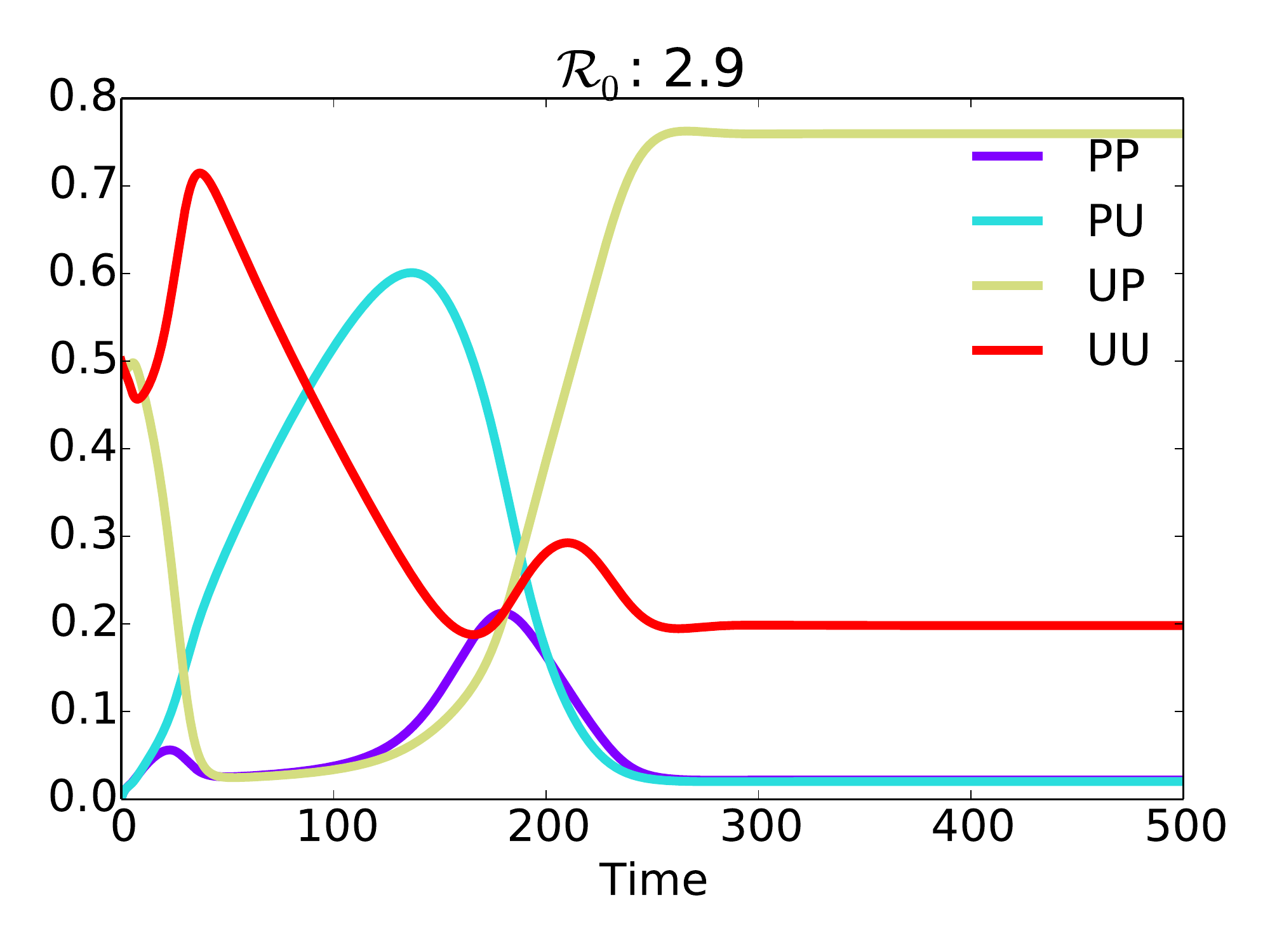}
		\caption{}	
	\end{subfigure}
	\begin{subfigure}[b]{0.475\textwidth}
		\includegraphics[width=\textwidth]{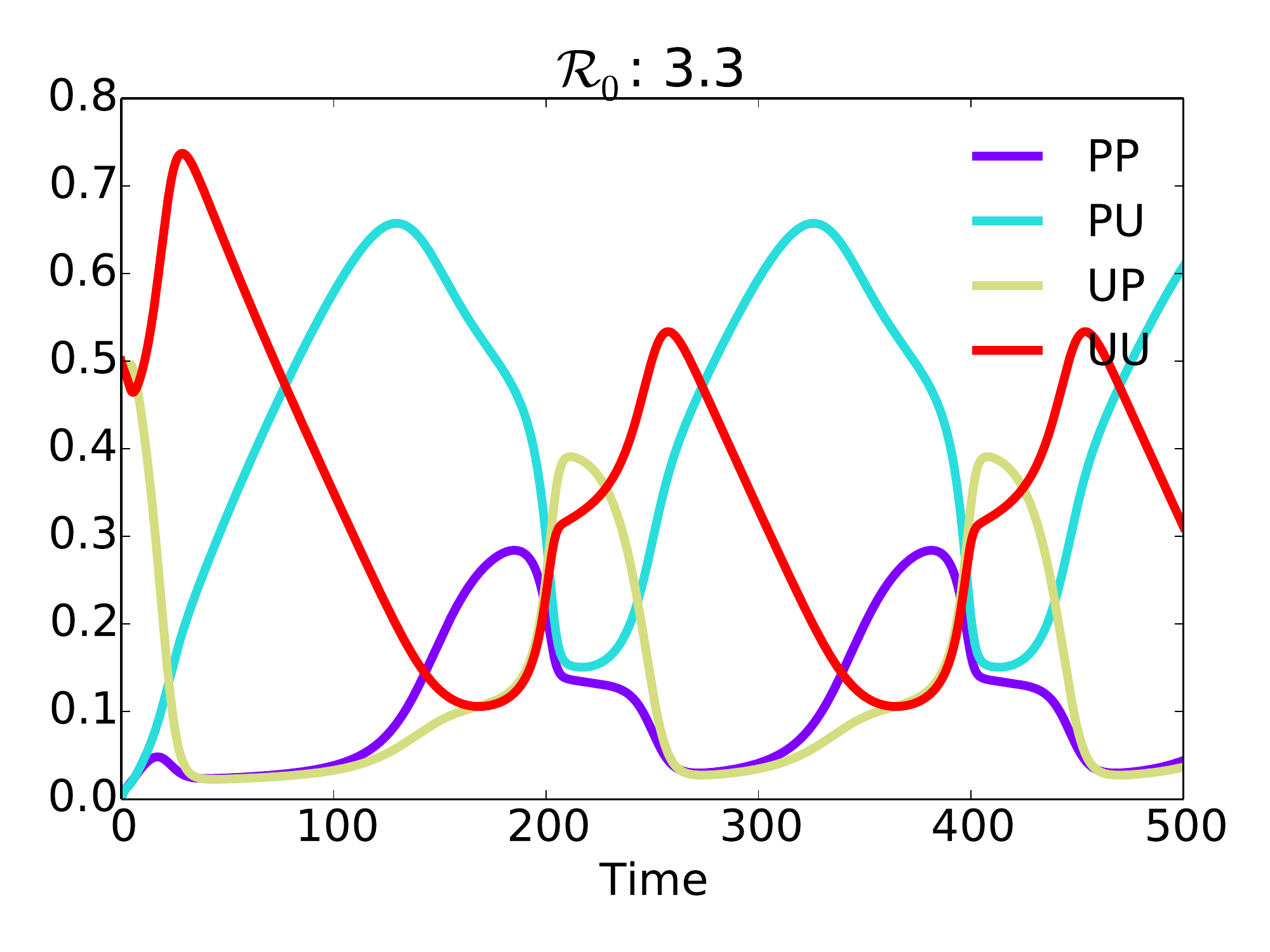}
		\caption{}	
	\end{subfigure}
	\begin{subfigure}[b]{0.475\textwidth}
		\includegraphics[width=\textwidth]{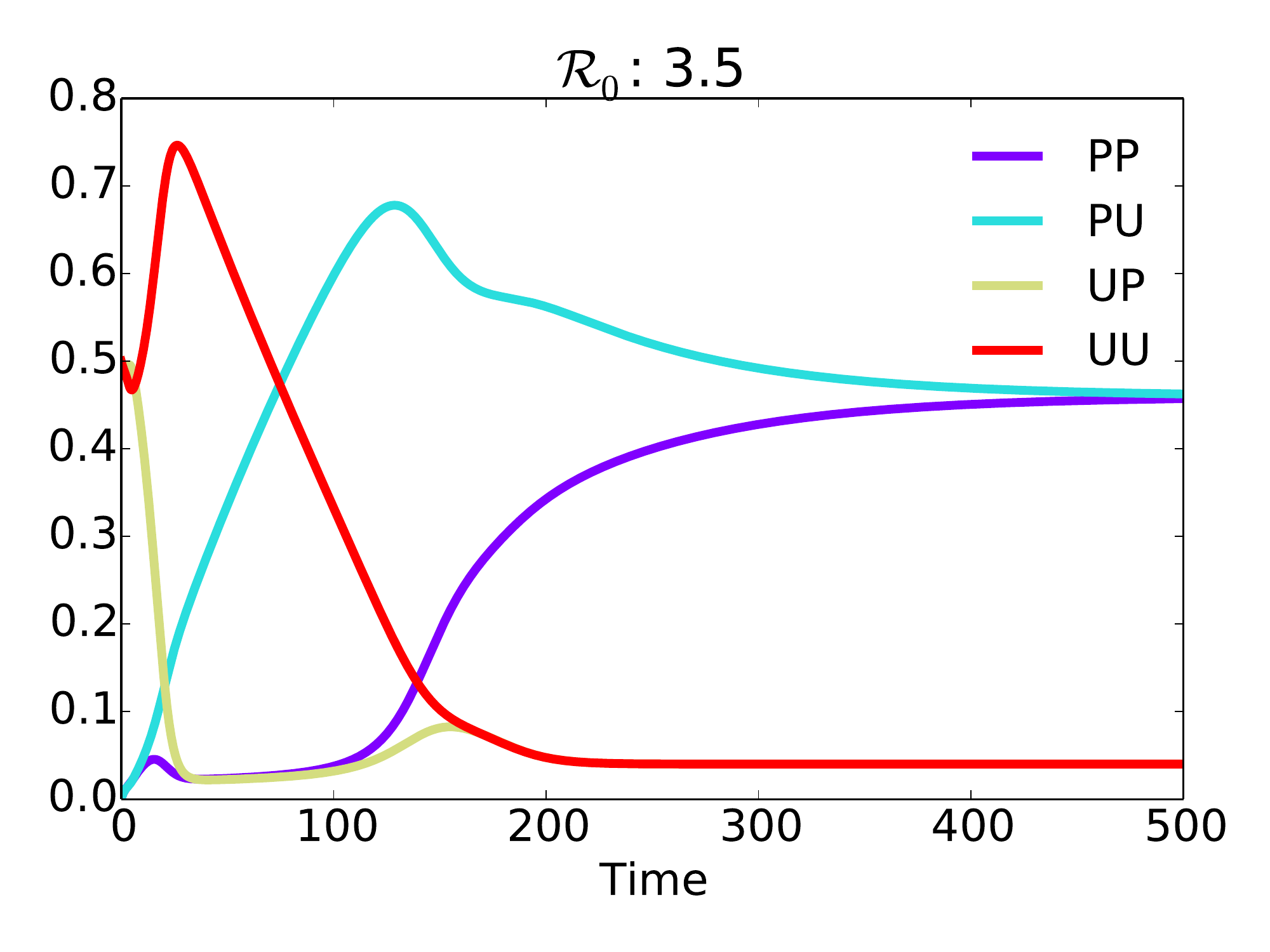}
		\caption{}	
	\end{subfigure}
	\caption{The distribution of type-contingent strategies as a function of time with $f(UU,0) = 0.5, \gamma = 0.5$ as in Figure \ref{fig::manytrajectories}.}
	\label{S1_Fig}
\end{figure}
\pagebreak

\section{Endemic Prevalence at $\Ro < 1$}
\begin{figure}[h]
	\centering
	\includegraphics[width=0.475\textwidth]{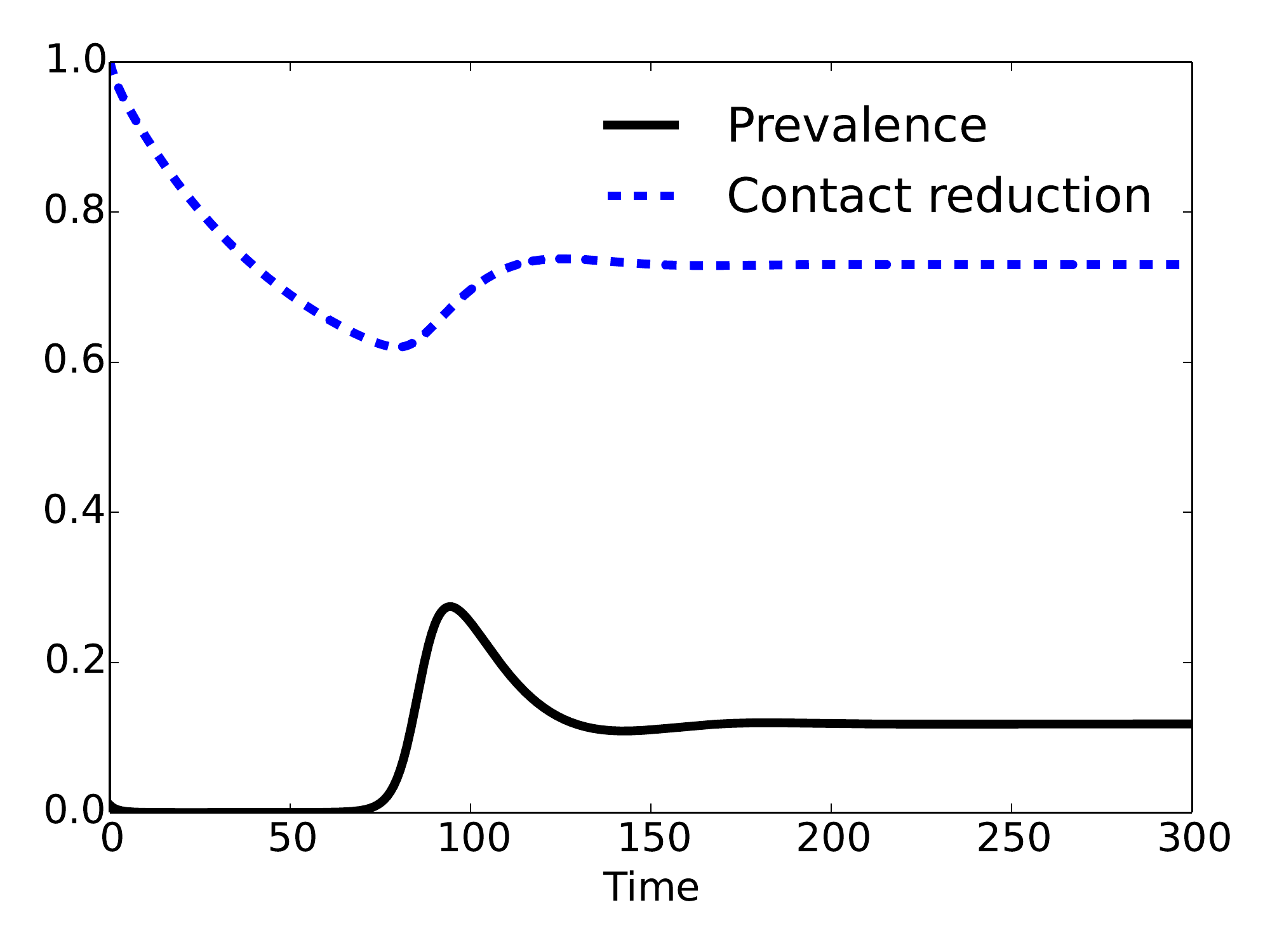}
	\caption{Prevalence and contact reduction from the combined model for \micolor{$\beta_b = 2.1, \gamma = 0.5, f(UU,0) = 0$.}}
	\label{S2_Fig}
\end{figure}

\begin{figure}[h]
	\centering
	\includegraphics[width=0.475\textwidth]{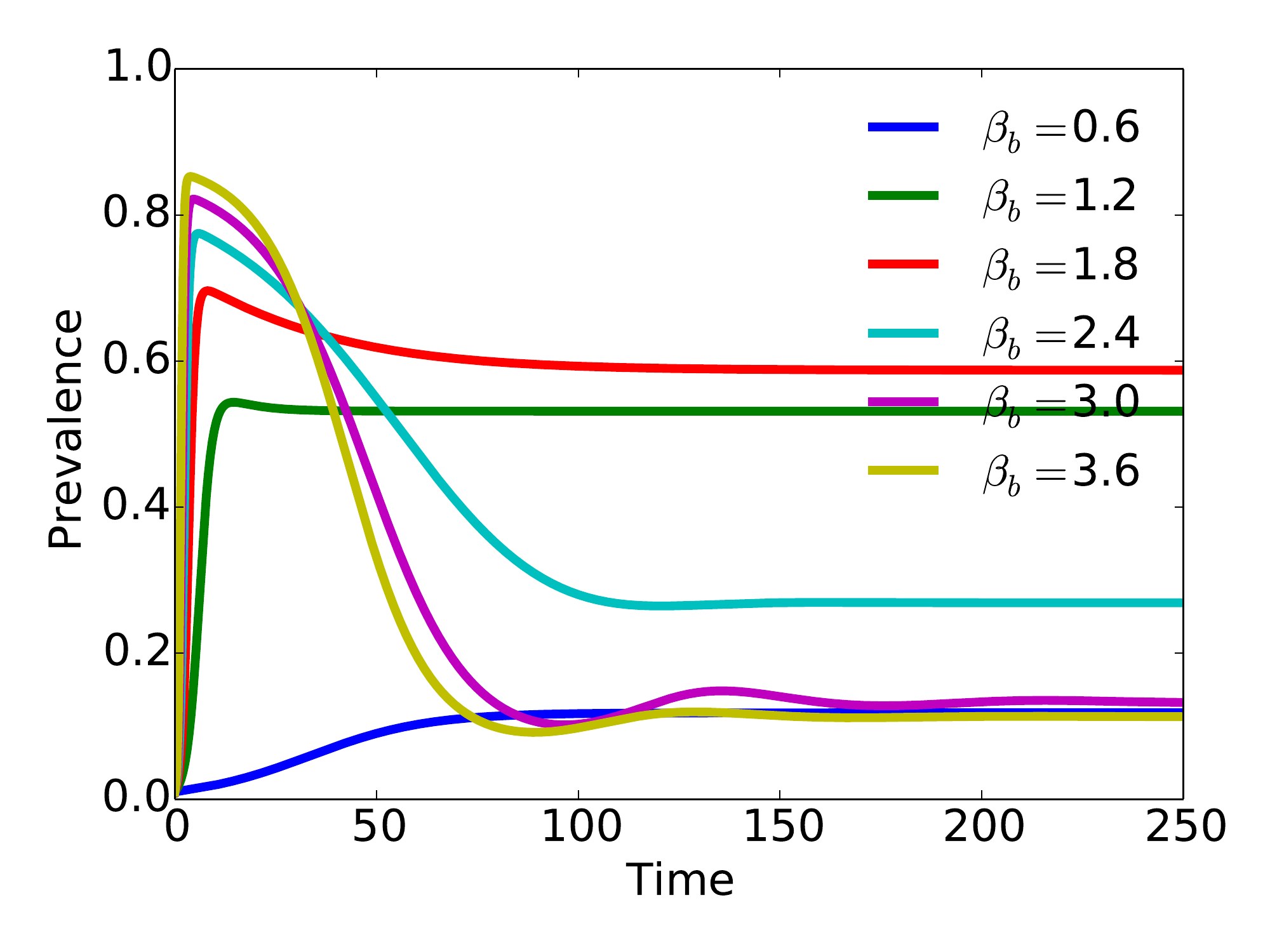}
	\caption{\micolor{Prevalence dynamics for the reduced model from Section \ref{sec::methods} where only susceptible individuals adapt their behavior.  For the simulations above $\gamma = 0.5$, $\mathbf{f}(0) = (0.0,1.0)$.}}
	\label{S4_Fig1}
\end{figure}

\begin{figure}[h]
	\centering
	\begin{subfigure}[b]{0.475\textwidth}
		\includegraphics[width=\textwidth]{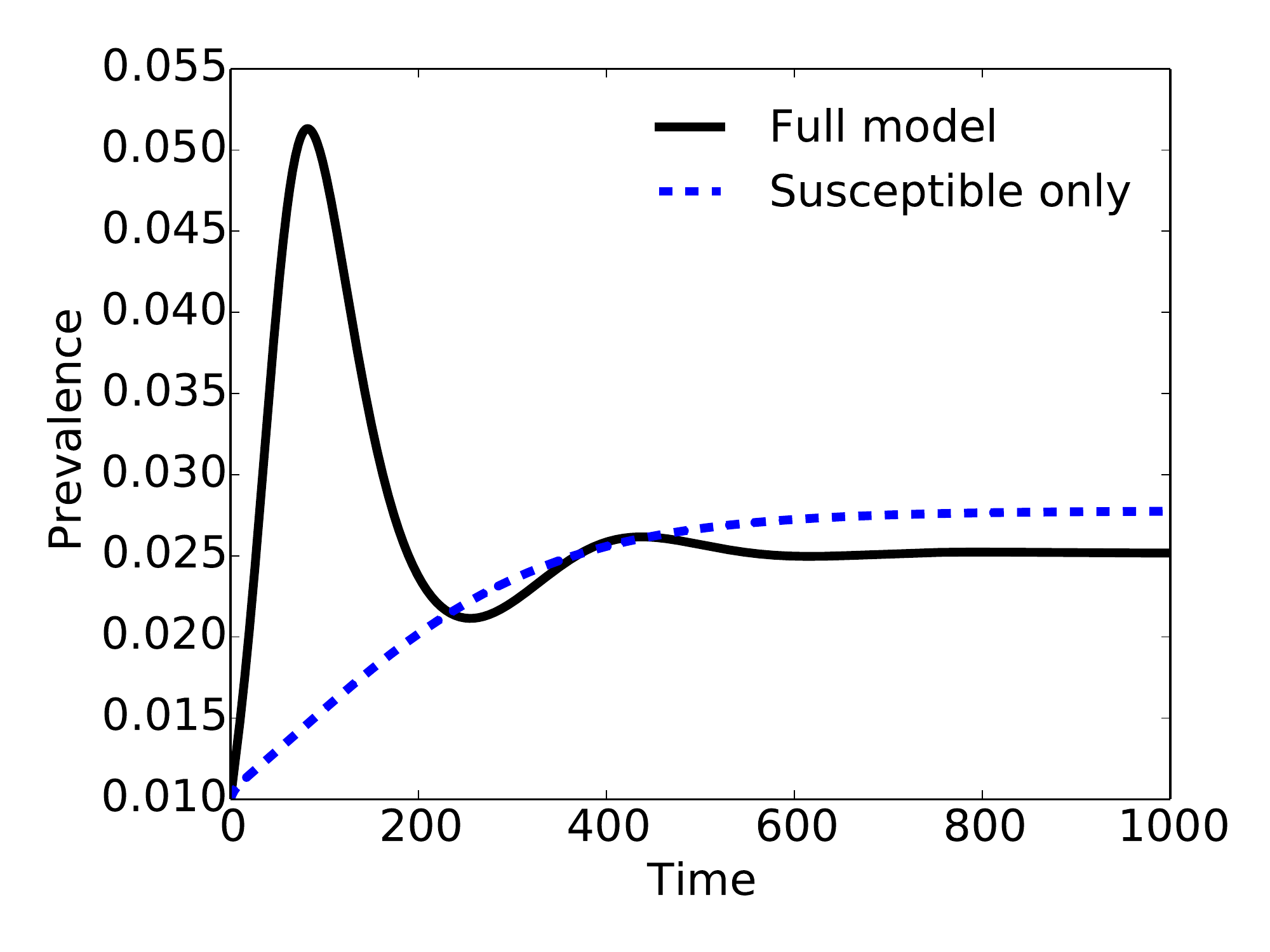}
		\caption{}
	\end{subfigure}
	\begin{subfigure}[b]{0.475\textwidth}
		\includegraphics[width=\textwidth]{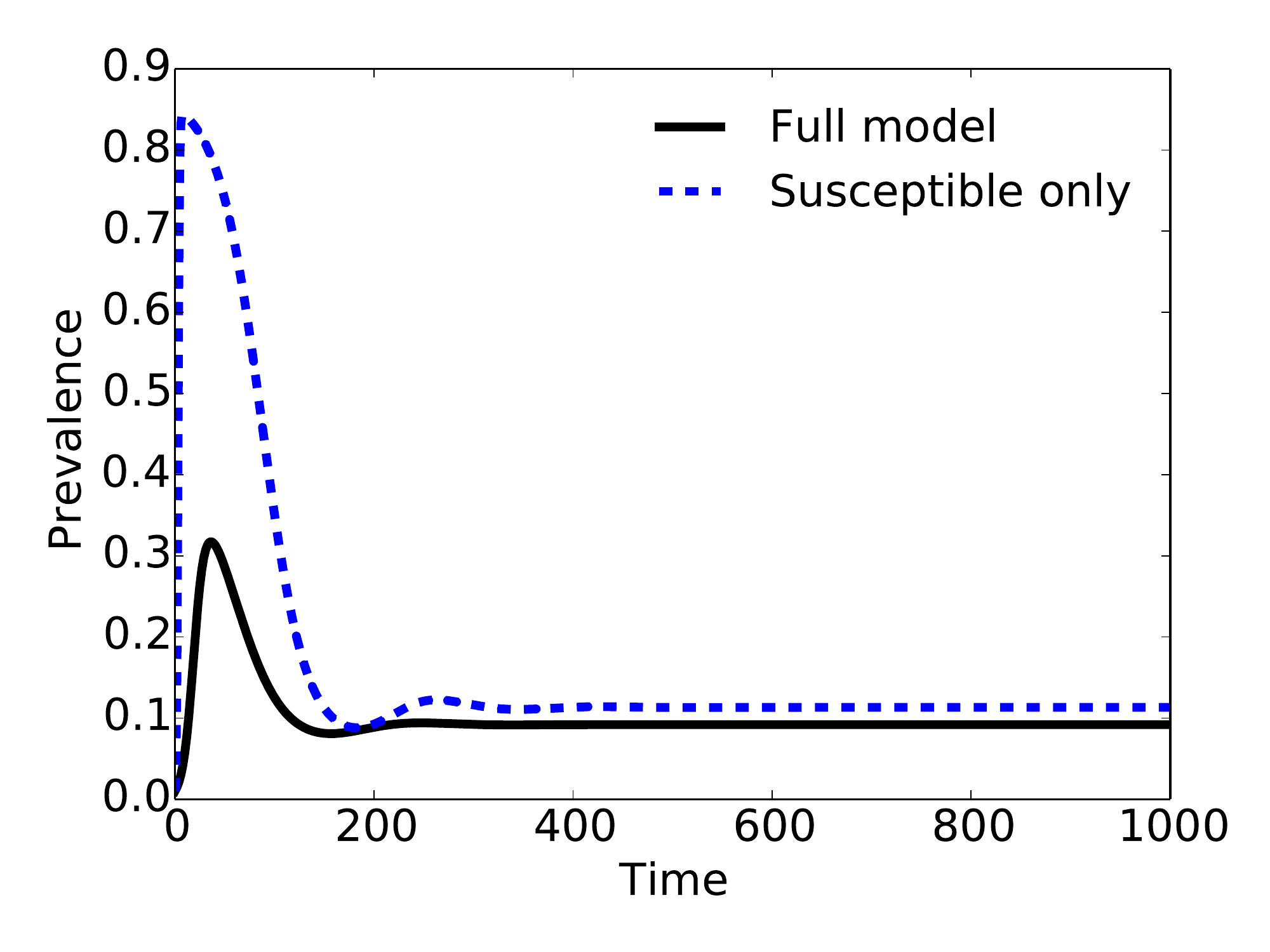}
		\caption{}
	\end{subfigure}
	\begin{subfigure}[b]{0.475\textwidth}
		\includegraphics[width=\textwidth]{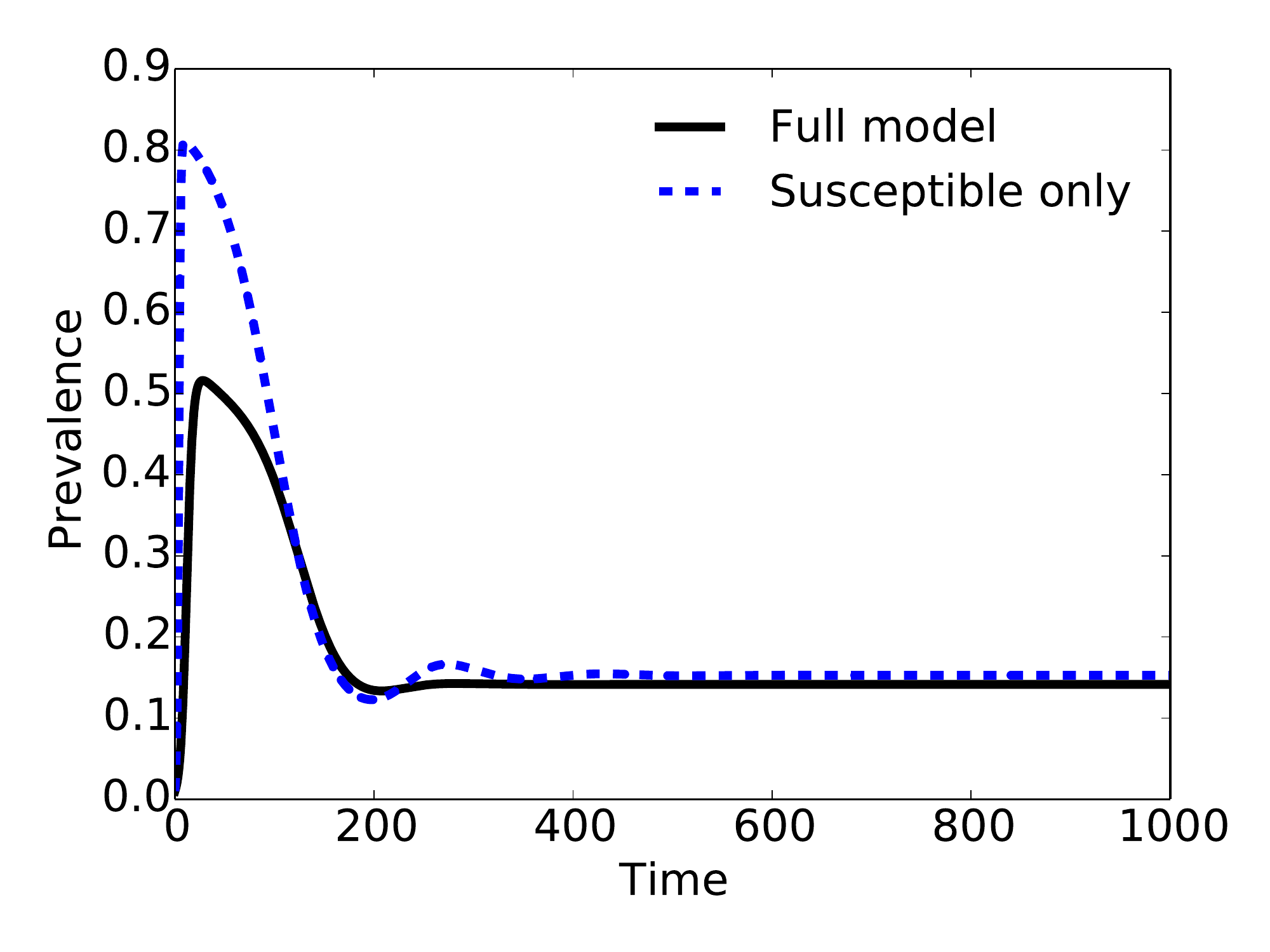}
		\caption{}
	\end{subfigure}
	\begin{subfigure}[b]{0.475\textwidth}
		\includegraphics[width=\textwidth]{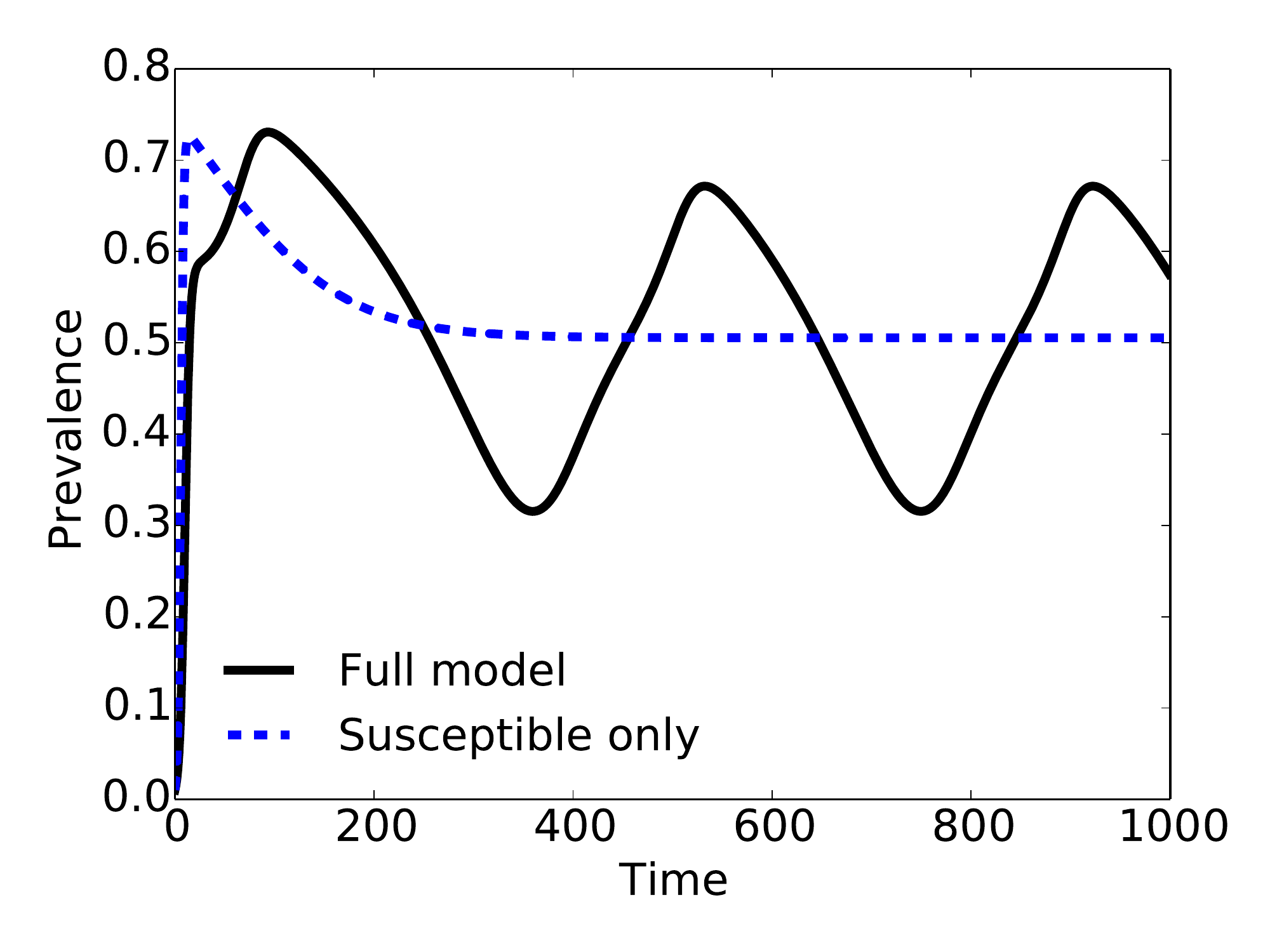}
		\caption{}
	\end{subfigure}
	\begin{subfigure}[b]{0.475\textwidth}
		\includegraphics[width=\textwidth]{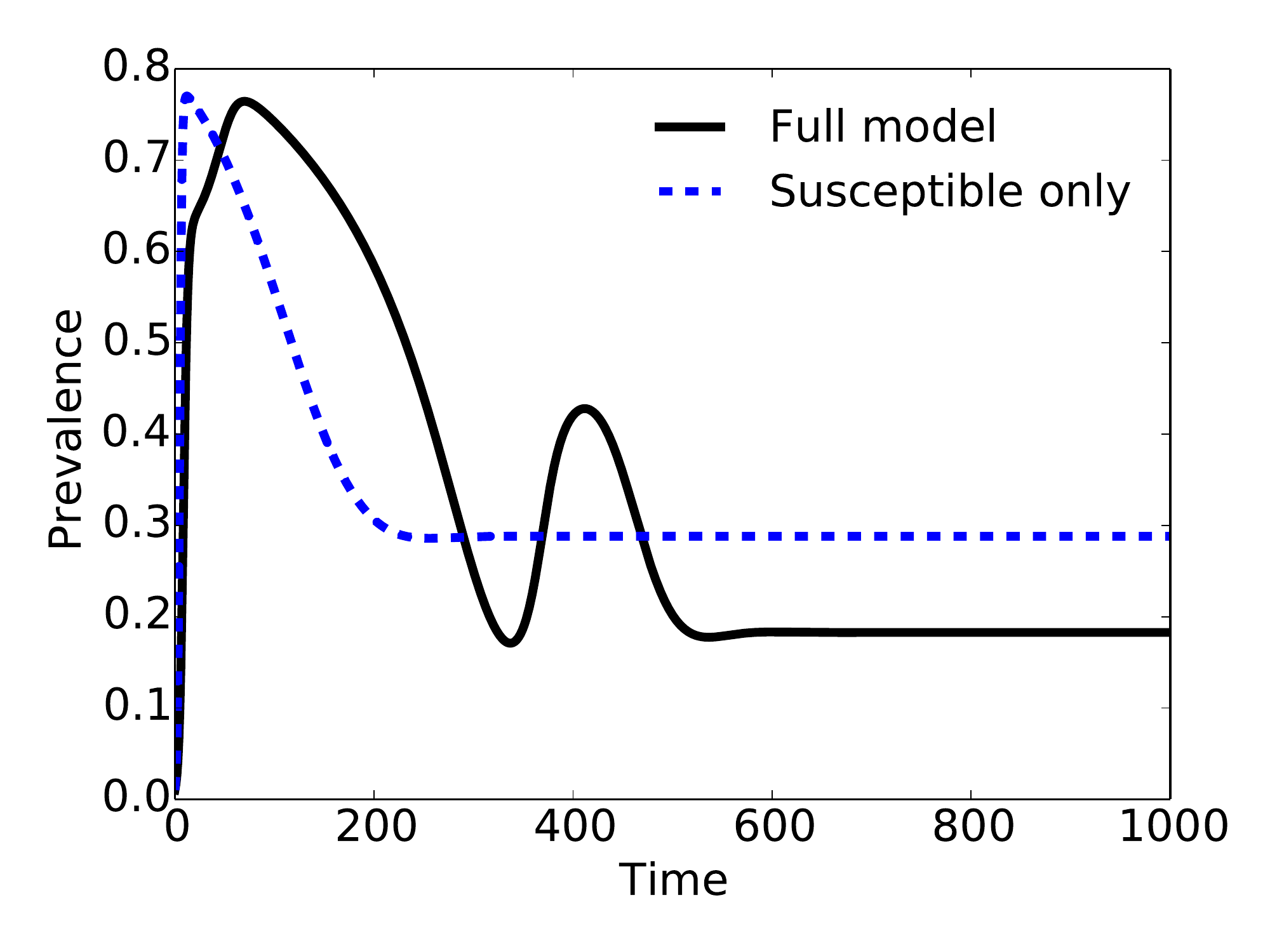}
		\caption{}
	\end{subfigure}
	\begin{subfigure}[b]{0.475\textwidth}
		\includegraphics[width=\textwidth]{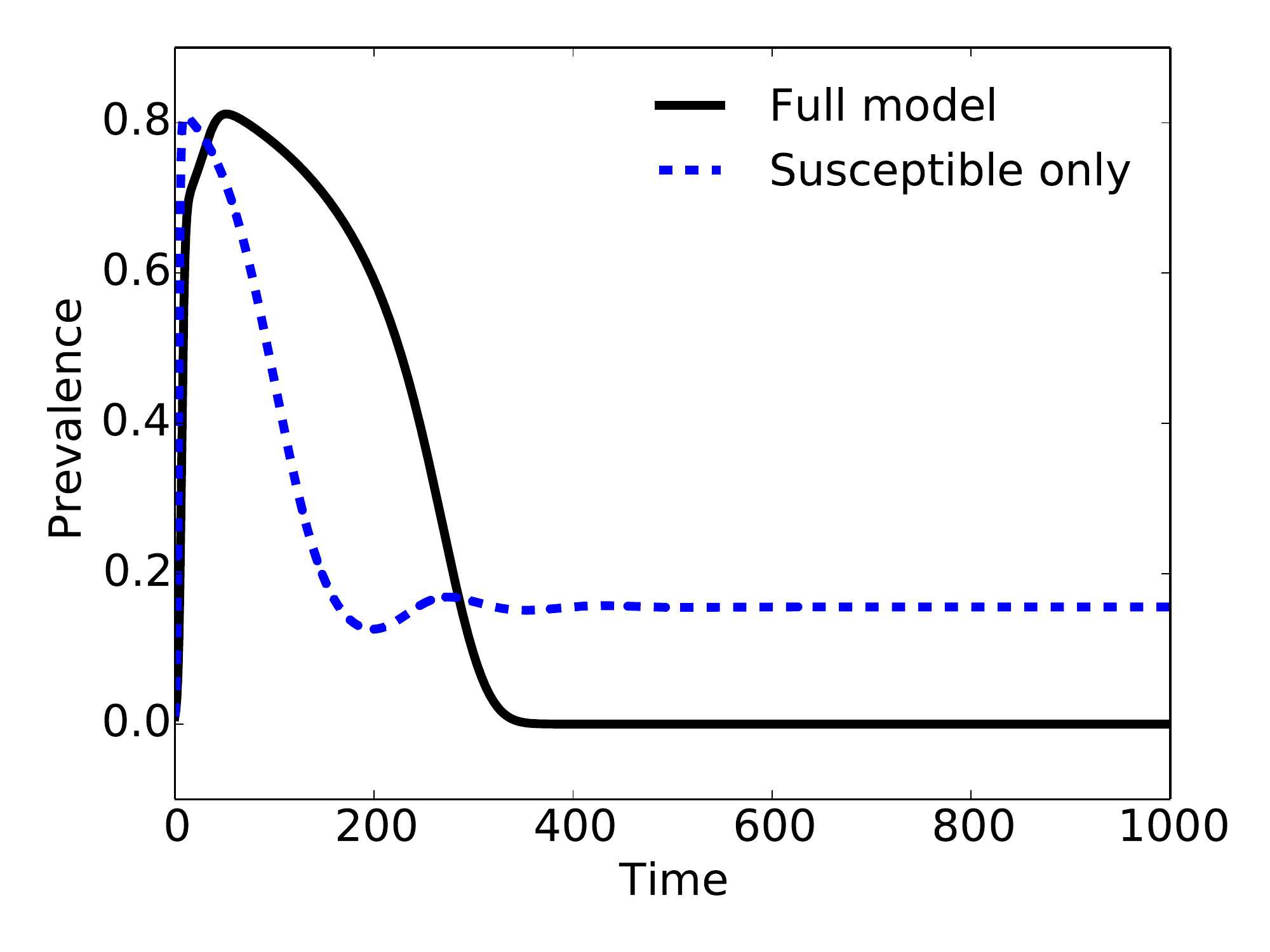}
		\caption{}
	\end{subfigure}
	\caption{\micolor{The reduced model of susceptible-only behavior change fit to simulated trajectories from the full model with $\gamma = 0.5$,  (a) Full model $\beta_b = 1.2$, best fit $\beta_b^*=0.542$, (b) full model $\beta_b = 1.8$, best fit $\beta_b^*=3.41$, (c) full model $\beta_b = 2.4$, best fit $\beta_b^* = 2.84$, (d) full model $\beta_b = 2.7$, best fit $\beta_b^*= 1.99$, (e) full model $\beta_b = 3$, best fit $\beta_b^* = 2.36$, (f) full model $\beta_b = 3.6$, best fit $\beta_b^* = 2.82$.}}
	\label{S4_Fig2}
\end{figure}

\micolor{
\section{Alternate Models}
\label{app::altmodels}
\subsection{SIRS Model}
\begin{equation}
\begin{aligned}
	\dot{S} &= \delta R - \beta SI \\
	\dot{I} &= \beta SI - \gamma I \\
	\dot{R} &= \gamma I - \delta R
\end{aligned}
\end{equation}
}
\micolor{
\subsection{Susceptible-only Behavior Change}
For the reduced model, we use the fitness functions
\begin{equation}
\begin{aligned}
	\phi(P_S,t) &=& Pr(S)[f(P_S,t)u_1(P_S,P_S,S,S) +f(U_S,t)u_1(P_S,U_S,S,S)]  \\
	&&+ Pr(I)[f(P_I)u_1(P_S,P_I,S,I) + f(U_I)u_1(P_S,U_I,S,I)] \\
	\phi(U_S,t) &=& Pr(S)[f(P_S,t)u_1(U_S,P_S,S,S) +f(U_S,t)u_1(U_S,U_S,S,S)]  \\
	&&+ Pr(I)[f(P_I)u_1(U_S,P_I,S,I) + f(U_I)u_1(U_S,U_I,S,I)]
\end{aligned}
\end{equation}
}
\micolor{
where the notation $P_S$ denotes a susceptible player choosing P (similarly for $U_S, P_I, U_I$) and $f(U_I)$ and $F(P_I)$ are fixed over time.  Since only susceptible players change their strategy, $\mathbf{f}(t) = (f(P_S,t),f(U_S,t))$, the mutation matrix for this game is $2\times2$ and we can use the two-dimensional system}
\micolor{
\begin{equation}
\begin{aligned}
	\dot{I} &= -\beta_b S_U f(U_I)I + \gamma I\\
	\dot{f}(U_S) &= q_{U_S P_S} \phi(P_S,t) f(P_S,t) + q_{U_S,U_S} \phi(U_S,t) f(U_S,t) - \bar{\phi} f(U_S,t)
\end{aligned}
\end{equation}
where $S_U = Sf(U_S,t)$.}
\micolor{
Figure \ref{S4_Fig1} shows prevalence trajectories for this model at increasing baseline effective contact rates.  Figure \ref{S4_Fig2} shows the best fit trajectories and parameter values for the reduced model compared to the full model.
}

\begin{figure}[h]
	\centering
	\begin{subfigure}[b]{0.475\textwidth}
		\includegraphics[width=\textwidth]{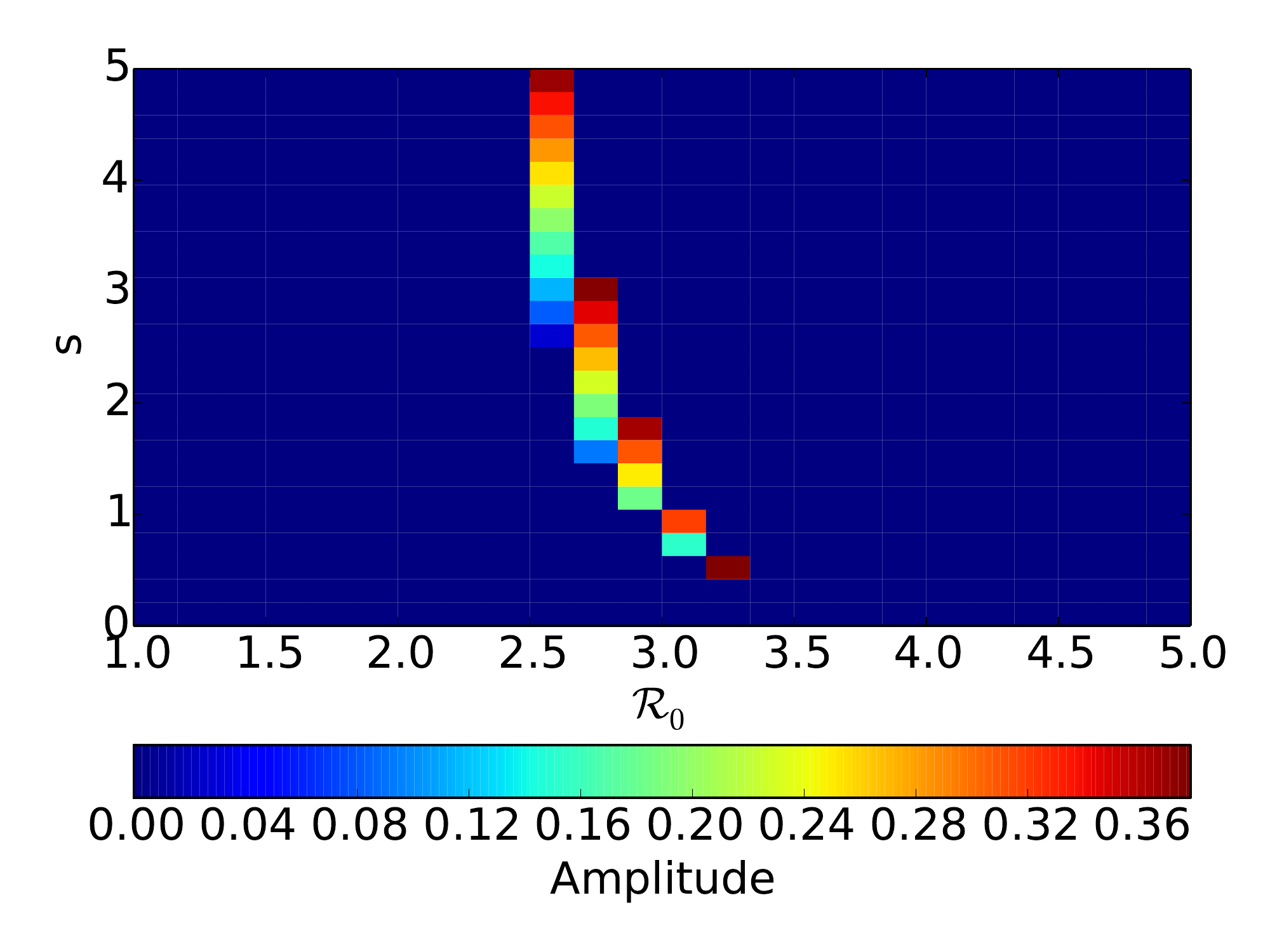}
		\caption{}
	\end{subfigure}
	\begin{subfigure}[b]{0.475\textwidth}
		\includegraphics[width=\textwidth]{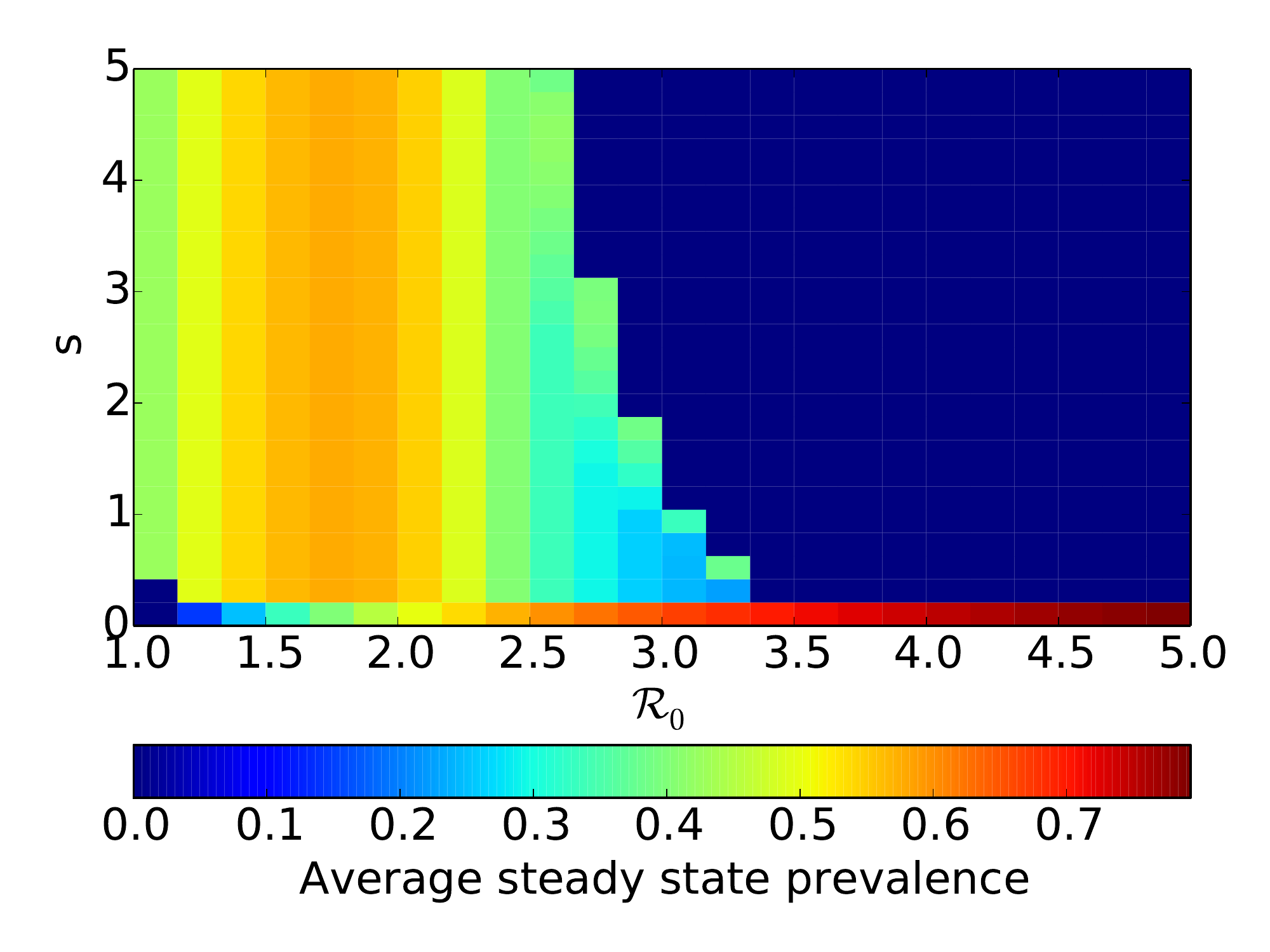}
		\caption{}
	\end{subfigure}
	\caption{\micolor{Long term dynamics of the combined model where infected individuals always prefer unprotected sex for increasing values of $\Ro$ and the behavioral scale parameter $s$.  (a) The amplitude of steady state prevalence oscillations.  (b) The average prevalence at steady state.}}
	\label{S5_Fig}
\end{figure}

\micolor{
\section{Alternate preferences}
\label{app::altcase}
We examined a case where infected individuals prefer unprotected sex over protected sex regardless of partner type, a phenomenon that has been observed empirically \cite{wenger1994sexual}.  For susceptible individuals, the type-dependent payoffs from the alternate game are the same as in Figure \ref{fig::payoffs}.  Without loss of generality, for an infected-type player 1 paired with a susceptible-type player 2, the type-dependent payoff matrix is}
\begin{center}
\begin{game}{2}{2}[Player~1][Player~2]
	     & $P$ & $U$ \\
	 $P$ & $b$ & $c$ \\
	 $U$ & $c$ & $a$ \\
\end{game}
\end{center}
\micolor{
Figure \ref{S5_Fig} shows the long-term behavior of the alternate model for a range of $\Ro$ and $s$.  This model only exhibits one oscillatory region and a higher endemic prevalence than the original model.  However, extinction still occurs at high $\Ro$.
}








\end{document}